\documentclass[a4paper,fleqn]{cas-dc}

\usepackage[comma]{natbib}
\renewcommand{\cite}{\citep}

\usepackage[utf8]{inputenc} 

\usepackage{graphicx}
\usepackage{lipsum}  
\usepackage{url}
\usepackage{fancyvrb}
\usepackage{xcolor}
\usepackage{hyperref}

\usepackage{multirow}
\usepackage{makecell}
\usepackage{adjustbox}
\usepackage{tabularx}

\usepackage[inline]{enumitem}
\usepackage{amsfonts}

\usepackage{graphicx}
\usepackage{stfloats}   

\usepackage{epigraph}
\usepackage{listings}
\usepackage{amssymb} 
\newcommand{\yes}{\checkmark}   
\newcommand{\no}{}      
\usepackage{booktabs}
\usepackage{balance}

\usepackage{array}

\usepackage{caption}
\usepackage{subcaption}
\usepackage{cleveref}
\crefformat{section}{\S#2#1#3} 
\crefformat{subsection}{\S#2#1#3}
\crefformat{subsubsection}{\S#2#1#3}
\crefformat{appendix}{\S#2#1#3}

\usepackage[textsize=tiny]{todonotes}

\usepackage{xcolor}
\usepackage{fancyvrb}
\usepackage[most]{tcolorbox}

\tcbset{
  rqstyle/.style={
    colback=gray!15, colframe=black!50, fonttitle=\bfseries,
    coltitle=black, sharp corners
  },
  ansstyle/.style={
    colback=gray!5, colframe=black!30, fonttitle=\bfseries,
    coltitle=black, sharp corners
  }
}

\newtcolorbox{researchquestion}[2][]{rqstyle, title=Research Question~#2,#1}
\newtcolorbox{answer}[2][]{ansstyle, title=Answer to RQ#2,#1}

\usepackage[ruled,vlined,linesnumbered]{algorithm2e}
\usepackage{changepage}
\SetKw{KwBy}{by}
\SetKw{Push}{push}
\SetKw{Empty}{empty}
\SetKw{PushBack}{pushback}
\SetKw{Top}{top}
\SetKw{Pop}{pop}
\SetKwProg{Fn}{Function}{}{end}

\newcommand{\revision}[2]{{\color{red} #1}}

\usepackage{pifont}

\newcommand{\journalcontent}[1]{}

\AtBeginDocument{%
  \providecommand\BibTeX{{%
    \normalfont B\kern-0.5em{\scshape i\kern-0.25em b}\kern-0.8em\TeX}}}

\newcommand{\ourtool}{{\sc STAFF}}

\usepackage{enumitem}
\usepackage{tikz}

\newcommand*\circnum[1]{%
  \tikz[baseline=(char.base)]{
    \node[shape=circle, fill=black, text=white, inner sep=1pt, minimum size=9pt] (char) {#1};}}

\begin{document}
\let\WriteBookmarks\relax
\def\floatpagepagefraction{1}
\def\textpagefraction{.001}

\shorttitle{STAFF: Stateful Taint‑Assisted Full‑system Firmware Fuzzing}    

\shortauthors{Izzillo et al.}  

\title [mode = title]{STAFF: Stateful Taint‑Assisted Full‑system Firmware Fuzzing}  



%


\author[1, 2]{Alessio Izzillo}[type=editor,
      auid=000,
      bioid=1,
      orcid=0009-0006-5502-5514
]
\ead{izzillo@diag.uniroma1.it}
\author[1]{Riccardo Lazzeretti}[type=editor,
      auid=000,
      bioid=1,
      orcid=0000-0003-3835-9679
]
\ead{lazzeretti@diag.uniroma1.it}
\author[2]{Emilio Coppa}[type=editor,
      auid=000,
      bioid=1,
      orcid=0000-0002-8094-871X]
\ead{ecoppa@luiss.it}






\affiliation[1]{organization={Sapienza University of Rome},
            country={Italy}}
\affiliation[2]{organization={LUISS University},
            country={Italy}}










\begin{abstract}
Modern embedded Linux devices, such as routers, IP cameras, and IoT gateways, rely on complex software stacks where numerous daemons interact to provide services. Testing these devices is crucial from a security perspective since vendors often use custom closed- or open-source software without documenting releases and patches. Recent coverage-guided fuzzing solutions primarily test individual processes, ignoring deep dependencies between daemons and their persistent internal state.


This article presents \ourtool{}, a firmware fuzzing framework for discovering bugs in Linux-based firmware built around three key ideas:
(a) \emph{user-driven multi-request recording}, which monitors user interactions with emulated firmware to capture request sequences involving application-layer protocols (e.g., HTTP);
(b) \emph{intra- and inter-process dependency detection}, which uses whole-system taint analysis to track how input bytes influence user-space states, including files, sockets, and memory areas;
(c) \emph{protocol-aware taint-guided fuzzing}, which applies mutations to request sequences based on identified dependencies, exploiting multi-staged forkservers to efficiently checkpoint protocol states.

When evaluating \ourtool{} on 15 Linux-based firmware targets, it identifies 42 bugs involving multiple network requests and different firmware daemons, significantly outperforming existing state-of-the-art fuzzing solutions in both the number and reproducibility of discovered bugs.


\end{abstract}




\maketitle


\section{Introduction}

Internet of Things devices are increasingly integrated into daily life across essential domains including healthcare, smart cities, transportation, and agriculture~\cite{iot_ecso}. This has opened the market to numerous manufacturers, each adopting custom embedded software that typically combines closed, proprietary code with open source components.
Such embedded software is usually developed according to two primary architectural approaches~\cite{holt2014embedded}.
Monolithic implementations integrate all components into a single program running directly on hardware, offering efficient resource utilization but limited flexibility. Alternatively, systems can be built on general-purpose operating systems, such as Linux, which provide standardized interfaces and greater development flexibility, though with additional overhead.
%
%
Several modern IoT devices often favor the latter architectural approach, increasingly adopting general-purpose operating systems\footnote{See the market analysis by Grand View Research: \url{https://www.grandviewresearch.com/horizon/statistics/embedded-software-market/operating-system/general-purpose-operating-system-gpos/global}.}.
However, these systems often include dozens or hundreds of interdependent binaries and configuration files from diverse sources, significantly expanding the attack surface and making devices particularly vulnerable to software flaws and security breaches~\cite{redini2020karonte,IOT-ANALYSIS-SECRYPT24}.

Identifying and prioritizing vulnerabilities — via firmware analysis — is a necessary precondition to guarantee system security. Direct testing of physical devices provides the most accurate results; however, it presents three main critical challenges~\cite{broekman2003testing}.
First, devices must be acquired, making it difficult to test a large variety of devices. Second, runtime inspection capabilities during execution are limited since most manufacturers disable hardware debugging interfaces (e.g., JTAG), making it difficult to exploit non-black-box analysis techniques. Third, physical hardware makes it challenging to perform parallel experiments on the same devices without acquiring multiple units. For these reasons, the research community has explored approaches that emulate firmware images, which can be obtained directly from vendors~\cite{dlink_taiwan_tsd, tplink_download}, retrieved during the update procedure, or manually extracted from device memory storage.

Emulation of firmware images enables scalable static and dynamic analysis but introduces its own challenges, as accurate rehosting\footnote{Rehosting refers to running software on a different platform than the one originally intended by the vendor.} remains an open research problem~\cite{wright2021challenges}.
Nonetheless, on one hand, several works~\cite {zhou2021automatic}
explore how to automatically model unknown peripherals and other vendor-specific hardware components. On the other hand, various rehosting frameworks~\cite{maier2020basesafe,hernandez2022firmwire,produit2025hexagon} 
have focused their attention on specific devices and have shown how to effectively emulate their firmware images. In this article, we focus on Linux-based systems that can be adequately emulated with state-of-the-art rehosting frameworks~\cite{liu2021firmguide,jun2023greenhouse}.
In particular, consistent with prior related work~\cite{triforceafl,zheng2019firm,equafl},
we primarily consider Linux-based routers and IP cameras.

%
One software testing technique exploited in several state-of-the-art firmware testing work~\cite{yun2022fuzzing,feng2022detecting,zhang2024survey,asmita2025bare}
is software fuzzing~\cite{afl,fuzzing-book,fuzzing-survey}.
This technique builds on the idea of testing the software a large number of times using different randomly mutated benign inputs. To evaluate the utility of each mutated input, fuzzing checks for crashes and, in their absence, exploits code coverage to understand whether a mutated input induced a different code execution path and thus could be a good candidate for further mutations. Projects such as OSS-Fuzz~\cite{ossfuzz,ossfuzz-results}
from Google have demonstrated the potential of software fuzzing by finding thousands of bugs and vulnerabilities in prominent and complex open-source projects.

\begin{figure}
    \centering
    \includegraphics[width=\columnwidth]{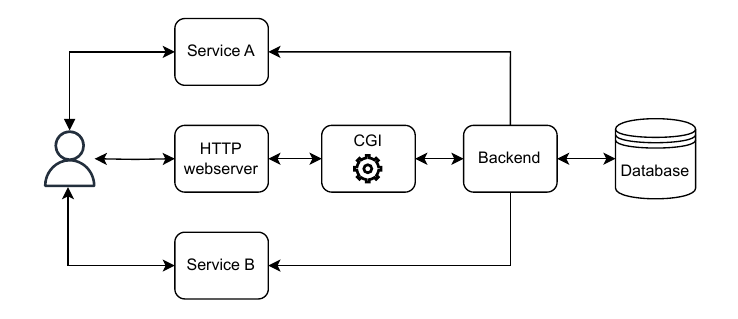}
    \caption{Example of a firmware exposing an HTTP webserver that internally interacts with a backend in charge of controlling other network services (e.g., DHCP and UPnP).}
    \label{fig:firmware_interactions}
\end{figure}

A major limitation of state-of-the-art work in firmware fuzzing is their limited analysis scope, as they fuzz the firmware by considering one service, i.e., one process, at a time. Furthermore, they see the process as a stateless application. However, a large number of devices are built atop stripped-down Linux distributions and expose a rich and complex surface of network-accessible services~\cite{tekeoglu2016testbed}. These include standardized protocols such as HTTP, DHCP, Telnet, FTP, UPnP, and SNMP, as well as vendor-specific APIs for configuration, telemetry, and remote management. The daemons behind these protocols are stateful, keeping track of their internal state across the expected sequences of requests. Furthermore, vendors build custom application protocols on top of these standardized protocols: for instance, as exemplified by Figure~\ref{fig:firmware_interactions}, a device may exploit HTTP as a bridge service that controls, configures, and impacts several other daemons. Such custom application protocols are not documented, with the exception of loose descriptions within user manuals, and involve several processes with internal states kept in memory or on the disk.

The tight integration of stateful protocols, long-lived user sessions, and persistent inter-process state makes systematic security testing of firmware images a significant challenge. Traditional greybox fuzzers, such as AFL~\cite{afl}, which have achieved significant success on traditional platforms, operate on single, stateless applications and cannot model inter-request dependencies. State-of-the-art firmware fuzzing approaches, including TriforceAFL~\cite{triforceafl},
FirmAFL~\cite{zheng2019firm},
and EQUAFL~\cite{equafl}
have primarily ported traditional fuzzers, focusing on the optimizations needed to amortize the substantial overhead introduced by emulation.
Prior work has considered applications implementing standard network protocols by exploring protocol-aware fuzzers. In particular, proposals such as AFLNet~\cite{pham2020aflnet}
and StateAFL~\cite{natella2022stateafl}
extend AFL's state model to cover common network protocols, but still focus on one daemon at a time and rely heavily on heuristics to infer valid transitions across internal states. Moreover, they target applications on traditional platforms and have not been ported to the context of firmware testing, where emulation overhead significantly impacts the execution cost during fuzzing.
Similarly, other fuzzers, such as VUzzer~\cite{rawat2017vuzzer},
Angora~\cite{chen2018angora},
WEIZZ~\cite{WEIZZ-ISSTA20},
and RedQueen~\cite{aschermann2019redqueen},
investigate how to exploit taint analysis, or lighter approximations of this technique, to understand how input bytes affect application execution. However, they target single applications. Full-system taint analysis has been explored in works such as DECAF++~\cite{davanian2019decaf++}
but has not been extensively used in the context of firmware fuzzing.




\paragraph{Our contribution.} In this article, we introduce \textbf{\ourtool{}}, \emph{Stateful Taint-Assisted Full-system Fuzzing}, a fuzzing framework aimed at firmware images exposing HTTP-based services that control and interact with multiple cooperating daemons. 
\ourtool{} encompasses three key stages:
\begin{enumerate*}[label=(\alph*)]
    
    \item {\em User-driven multi-request recording} (Section~\ref{ssec:exploration_corpus_gathering}),
    which closely monitors user interactions with the emulated firmware to capture rich but minimal request sequences involving an HTTP-based application-layer service that controls and interacts with a potentially large number of daemons. Prior work often undermines the importance and availability of meaningful request sequences. While \ourtool{} cannot make this phase fully automatic, it at least makes it a piece of the puzzle, providing support to users interested in fuzzing a firmware image.

    \item {\em Intra- and inter-service dependency detection} (Section~\ref{ssec:dataflow_analysis}),
    which exploits whole-system taint analysis to track how input bytes influence user-space states, including files, sockets, and memory regions, while keeping track of the request sequences possibly needed to reach interesting execution states. 

    \item {\em Protocol-aware taint-guided full-system fuzzing} (Section~\ref{ssec:fuzzing}),
    which exploits the dependencies identified during whole-system taint analysis to effectively devise mutations to request sequences during fuzzing. To make the testing process more efficient, \ourtool{} implements multi-staged forkservers to efficiently checkpoint complex protocol states.
  \end{enumerate*}

In more detail, our contributions can be summarized as follows:

\begin{itemize}

  \item We present the most significant fuzzing literature relevant to our context, explaining the key ideas behind them (Section~\ref{sec:background}) while also pinpointing their limitations when put in perspective on a real-world case study (Section~\ref{sec:motivation}).

  \item We introduce our firmware fuzzing framework \ourtool{} (Section~\ref{sec:approach}), which encompasses the three key stages outlined above:
  \begin{enumerate*}[label=(\alph*)]
    
    \item {\em User-driven multi-request recording}; 

    \item {\em Intra- and inter-service dependency detection}; 

    \item {\em Protocol-aware taint-guided full-system fuzzing}. 
  \end{enumerate*}

  \item We describe the essential implementation details behind our framework (Section~\ref{sec:implementation}). \ourtool{} extends and improves
  DECAF++
  for the whole-system taint analysis.
  We also present several optimizations required to make \ourtool{} more effective when considering real-world firmware images.

  \item We discuss the results of an experimental evaluation (Section~\ref{sec:evaluation}) where \ourtool{} is used on 15 firmware images, comparing its effectiveness with system-mode TriforceAFL
  and system-mode AFLNet$^{\dagger}$\footnote{We ported the ideas behind AFLNet into a multi-process firmware fuzzing framework dubbed AFLNet$^{\dagger}$.}.
  \ourtool{}  identifies 42 bugs involving multiple network requests and different firmware daemons, significantly outperforming existing state-of-the-art fuzzing solutions in both the number and reproducibility of discovered bugs.

\end{itemize}

To encourage further research, we make\footnote{We plan to release the code after acceptance.} our contributions available at {\tt \url{https://github.com/alessioizzillo/STAFF}}.

\section{Background}
\label{sec:background}

In this section, we provide the key ideas behind firmware fuzzing, presenting the most relevant state-of-the-art work that we consider throughout the article.

\subsection{Software Fuzzing}
\label{ssec:software_fuzzing}

\begin{figure}[!t]
\centering
\includegraphics[height=5cm]{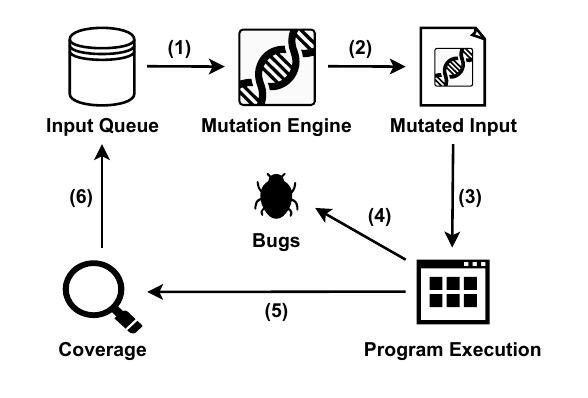}
\caption{Workflow of a coverage-guided fuzzer.}
\label{fig:fuzzing-workflow}
\end{figure}

Software fuzzing is a dynamic testing technique~\cite{fuzzing-survey,godefroid2020fuzzing,fuzzing-book} that generates randomly mutated inputs and fed them to an application with the ultimate goal of finding bugs and vulnerabilities. In this article, we consider coverage-guided fuzzing, a flavor that has gained widespread adoption through popular fuzzing frameworks such as AFL (American Fuzzy Lop)~\cite{afl} and libFuzzer~\cite{libfuzzer}, both recognized for their effectiveness in large-scale vulnerability discovery~\cite{ossfuzz,ossfuzz-results}. 
%
%
%
The main steps of a coverage-guided fuzzer, as illustrated in Figure~\ref{fig:fuzzing-workflow}, can be outlined as follows:

\setlist[enumerate]{itemsep=1pt, topsep=3pt}
\begin{enumerate}
    \item \emph{Input Selection:} The fuzzer selects a test case from the \textit{input queue}, a prioritized collection of candidate inputs. This queue is initially seeded with inputs, dubbed \textit{seed corpus}, provided manually by the user or extracted from existing data samples. The selection process may incorporate various heuristics to prioritize inputs that are likely to exercise unexplored code paths or trigger new program behaviors.

    \item \emph{Input Mutation:} The selected input is subjected to one or more mutation strategies to produce a new test case. These mutations are typically lightweight and computationally inexpensive operations such as bit flipping, arithmetic adjustments, insertion of random byte sequences, injection of semantically meaningful constants (e.g., platform-specific magic numbers or dictionary tokens), or recombination with other inputs in the queue. The mutation scheme and intensity are often stochastically determined to maintain diversity in the generated inputs.

    \item \emph{Program Execution and Monitoring:} The target software is executed with the mutated input. The fuzzer actively monitors the execution for anomalous behaviors, including crashes, timeouts, or hangs. Meanwhile, code coverage information is collected to evaluate the input impact on the application behavior. This instrumentation can be achieved either statically at compile time, when the source code is available, or dynamically at runtime using techniques such as dynamic binary translation (e.g., via QEMU~\cite{qemu}).

    \item \emph{Crash Handling and Deduplication:} If the execution results in a crash, the corresponding input is preserved in a dedicated \textit{crash queue}.
    To avoid redundant inputs, a crash deduplication mechanism is employed to identify and discard semantically equivalent failures. 

    \item \emph{Code Coverage Feedback:} Inputs that do not cause crashes may still contribute if they increase the overall code coverage. A coverage-guided fuzzer leverages fine-grained execution metrics to determine whether an input has led to the exploration of previously unexecuted control flow paths or exercised rare branches. Such inputs are considered \textit{interesting}.

    \item \emph{Queue Augmentation:} If an input is deemed interesting based on its contribution to program coverage or behavioral diversity, it is reinserted into the input queue for further mutation in future fuzzing iterations.

\end{enumerate}

This iterative \textit{mutate-and-execute} paradigm is repeated at scale, often millions or billions of times, enabling the fuzzer to autonomously discover deep vulnerabilities and edge-case behaviors even in complex applications.

\subsection{Firmware Fuzzing}

Given the impressive results obtained by coverage-guided fuzzing on traditional platforms, researchers have naturally explored these techniques for firmware analysis~\cite{gao2022fw,shen2022tardis,FUZZPLANNER-VIZSEC23}.
Two primary approaches have emerged for fuzzing firmware images, each presenting distinct trade-offs between accuracy and effectiveness.

The first approach~\cite{triforceafl} employs full-system emulation, where the entire firmware image runs in an emulated environment while code coverage is monitored either system-wide
or for specific processes when targeting individual applications. While this strategy provides high fidelity by preserving the complete execution context, it suffers from significant execution overhead. System emulation is slow, substantially limiting the fuzzing throughput and potentially compromising the overall effectiveness of the testing process.

The second approach, adopted by several fuzzing frameworks~\cite{zheng2019firm,equafl}, attempts to overcome these performance limitations through a hybrid emulation strategy. These tools initially use system emulation to bootstrap the firmware into a stable running state, then migrate target applications to faster user-space emulators. This migration can dramatically increase execution speed and fuzzing attempts. However, the approach requires switching back to system-mode emulation whenever the application interacts with the kernel through system calls, introducing both complexity and overhead.

This hybrid strategy faces several fundamental limitations. The performance benefits diminish as kernel interactions increase, since the costly synchronization and context switching between emulation modes becomes a bottleneck. More critically, this approach is fundamentally designed for isolated application testing, which can poorly match the reality of modern firmware architectures. Modern firmware images typically feature complex ecosystems of interdependent daemons that communicate through various inter-process communication mechanisms. User-space emulation struggles to accurately capture these inter-service dependencies, potentially missing critical bugs that arise from component interactions rather than isolated application logic.

In this paper, our goal is to simultaneously test multiple cooperating services within a firmware. To achieve this goal, we adopt fuzzing strategies based on system-mode emulation, which enables comprehensive analysis of inter-service interactions.  Due to the huge overhead from the emulation, which can drastically reduce the number of fuzzing attempts, it is pivotal to generate inputs based on targeted mutations guided by identified intra- and inter-process dependencies, thus hopefully maximizing the chances of exploring interesting behaviors across different processes.

\subsection{Protocol-Aware Fuzzing}
\label{ssec:protocol-fuzzing}

Different works
have considered fuzzing of applications implementing network protocols. A prominent example is AFLNet~\cite{pham2020aflnet},
which extends the popular AFL~\cite{afl} in order to cope with sequences of requests instead of handling the application input as a single monolithic file. We now presents the main key ideas behind AFLNet since they are relevant for the remainder of this article.

\paragraph{Input as ordered regions.} For applications implementing network protocols, a valid input interaction may depend not only on a single request but on ordered sequences of requests, dubbed by AFLNet as \textit{regions}, where each region corresponds directly to a single protocol request. While classic AFL treats inputs as a single, monolithic file, AFLNet interprets them as multi-step session governed by an implicit protocol state machine.
For instance, consider the requests:
\begin{enumerate}[label=\protect\circnum{\arabic*}]
  \item {\em Initial page request}:
\vspace{-2mm}
\begin{small}
\begin{verbatim}
GET /index.html HTTP/1.1
Host: example.com
\end{verbatim}
\end{small}
\vspace{-2mm}
  \item {\em Login request}:
\begin{small}
\vspace{-1mm}
\begin{verbatim}
POST /login HTTP/1.1
Host: example.com
Content-Length: 30
\end{verbatim}
\vspace{-3mm}
\begin{verbatim}
username=admin&password=admin
\end{verbatim}
\end{small}
\vspace{-2mm}
  \item {\em Authenticated dashboard request}:
\vspace{-2mm}
\begin{small}
\begin{verbatim}
GET /admin/dashboard HTTP/1.1
Host: example.com
\end{verbatim}
\end{small}
\vspace{-1mm}
\end{enumerate}

\noindent Each HTTP request (\texttt{GET}, \texttt{POST}, etc.) is treated as a distinct \textit{region}. AFLNet extracts these regions from captured network traces (e.g., PCAP files) and mutates them individually while preserving their order. This notion of \textit{region-aware mutation} allows AFLNet to maintain valid session structures: instead of corrupting random bytes, it mutates meaningful protocol elements, such as HTTP methods (\texttt{GET} $\rightarrow$ \texttt{PUT}), URIs, or header values.
Through region-aware mutation, AFLNet ensures that mutated regions are still likely to pass initial protocol checks, increasing the chances of reaching deeper code paths such as login logic or administrative endpoints.


\paragraph{Implicit Protocol State Machine (IPSM).} While mutating sequences, AFLNet observes the server responses, typically in the form of HTTP status codes (e.g., \texttt{200 OK} or \texttt{401 Unauthorized}). By correlating these codes with request sequences, AFLNet dynamically builds an \textit{Intermediate Protocol State Machine}, where states correspond to unique server response codes or combinations of codes, while transitions represent HTTP requests (regions) that move the server from one state to another.
For example:
\[
\textit{Initial} \xrightarrow{\texttt{GET /index}} \textit{Unauthenticated} \xrightarrow{\texttt{POST /login}} \textit{Authenticated}
\]

\noindent This protocol state machine is inferred dynamically, without requiring a formal protocol specification. As AFLNet explores more sequences, the protocol state machine becomes richer, capturing both expected and unexpected states (e.g., unexpected error responses like \texttt{500 Internal Server Error}).

\paragraph{Target State Selection.}
%
Not all protocol states are equally valuable for exploration. AFLNet introduces target state selection heuristics to focus on the most promising parts of the protocol state machine. These heuristics consider: rarity, states that are reached infrequently, states historically linked to new code paths or crashes. 
Once a target state is selected, AFLNet chooses sequences from the corpus that reliably reach it. This ensures that mutations start from a meaningful context, such as an authenticated session, rather than always from the initial unauthenticated handshake.

\paragraph{State-Aware Mutations.}
Adopting the \emph{region} abstraction, let $M = \langle M_1, M_2, M_3 \rangle$ denote a full input interaction sequence decomposed into a prefix $M_1$ (the region or regions that reach that chosen state $s$), the target $M_2$ (the intermediate region or regions targeted for mutation), and a suffix $M_3$ (remaining region or regions that may observe or materialize the effects of $M_2$). The state-aware strategy mutates only the target $M_2$, producing:
\[
M' = \langle M_1, \mathrm{mutate}(M_2), M_3 \rangle.
\]
Preserving $M_1$ maintains the transition path to the target state so that the mutated target is exercised in the correct protocol context. Retaining $M_3$ ensures any downstream consumers, committers, or cleanup handlers that observe the mutated data still execute and can therefore expose faults that only manifest after the mutation is materialized. For example:
\begin{itemize}
    \item \textbf{Prefix $M_1$:} \texttt{GET /index} followed by \texttt{POST /login}, which reliably establishes an authenticated state.
    \item \textbf{Target $M_{2}$:} Configuration change requests, e.g., \texttt{POST /lan\_settings.cgi} with parameters for IP address, netmask, or DHCP range. These intermediate requests are mutated by the fuzzer to exercise parsing and validation logic.
    \item \textbf{Suffix $M_{3}$:} Commit or apply request, e.g., \texttt{POST /apply.cgi} or \texttt{POST /xmldb.php}, which finalizes the configuration changes by writing them to persistent storage and/or notifying backend daemons.
\end{itemize}

This \textit{state-aware approach} significantly improves code coverage compared to stateless fuzzing or blindly mutating concatenated HTTP sessions.


\paragraph{Timeout-based delimitation.}  
AFLNet determines response boundaries using two types of timeouts at the socket layer: a \emph{polling timeout}, which specifies the maximum period to wait before concluding that no further responses will arrive, and a \emph{socket timeout}, which defines the wait time for each individual response. These values act as heuristics to delimit request-response pairs when explicit protocol boundaries are not available.

\subsection{Dynamic Taint Analysis}\label{sec:dynamic_taint_analysis}

Dynamic Taint analysis~\cite{newsome2005dynamic,clause2007dytan}
is a dynamic program analysis technique that tracks how sensitive input data propagates throughout program computation. Unlike symbolic execution~\cite{BCDDF-CSUR18,FUZZYSAT-ICSE21,SYMFUSION-ASE22,DEBUG-CE-JSS24}, dynamic taint analysis tracks data flow dependencies but does not capture the specific transformations or constraints applied to the data.
The main idea is that inputs are initially tainted when they are fetched from (untrusted) sources (e.g., read from a file) and then tracked during the execution to check whether they, or any derived data from them, reach sinks (e.g., security-sensitive operation). To propagate taints across variables, registers, and memory, the analysis applies propagation rules at each executed instruction, introducing significant execution overhead.

\paragraph{Taint-Guided Fuzzing.}
Given the success of taint analysis in different applications, such as the identification of information leakage
~\cite{yang2012leakminer},
several fuzzing works
have explored how to exploit this technique to devise targeted mutations~\cite{bekrar2012taint,liang2022pata}. For instance, VUzzer~\cite{rawat2017vuzzer}
exploits dynamic taint analysis to identify which input bytes may affect the decision points of a program, then exploits this knowledge during the mutation phase. However, a notable downside is the overhead introduced by dynamic taint analysis; hence, more recent works have investigated approximations of taint analysis that heuristically attempt to understand whether some input bytes have flowed into, e.g., a comparison operand. For instance, REDQUEEN~\cite{aschermann2019redqueen}
introduces the concept of {\em input-to-state correspondence} that leverages the fact that part of the input is often used as is, or after a few standard manipulations (e.g., addition by 1 or base64 encoding), in a comparison operation.

\paragraph{Whole System Taint Analysis.}
Dynamic taint analysis can be performed over a single application or on an entire system. The latter case is often called {\em whole-system taint analysis} and is particularly challenging to realize since {\em all} operations performed by the system must be tracked during execution, with sources and sinks that may be defined at software-hardware boundaries. In this article, we build our proposal on top of DECAF++~\cite{davanian2019decaf++},
a dynamic taint analysis framework for QEMU that embeds propagation rules for TCG, QEMU's architecture-independent intermediate representation.

\section{Motivation}
\label{sec:motivation}


Modern embedded devices may expose a variety of network protocols, such as HTTP, DHCP, ARP/ICMP, Telnet, FTP, UPnP, and SNMP, through different daemons that can often be configured and controlled via an HTTP interface. 
The device's internal architecture is far from simple. Manufacturers often combine custom, closed-source components with open-source components common to several Linux platforms, devising workflows that are undocumented and may contain bugs or vulnerabilities~\cite{chen2016towards}.
A common entry point for these vendor-specific workflows is the HTTP interface. Indeed, HTTP requests often serve as entry points to complex multi-process workflows, where Common Gateway Interface (CGI)
handlers written in PHP or other languages invoke internal backend daemons to read and write persistent configuration state, as shown in~\autoref{fig:firmware_interactions}. This layered design means that vulnerabilities may arise not from a single malformed HTTP request alone but from a \emph{sequence} of interdependent requests that transit through and affect multiple processes.

\paragraph{Motivating Case Study.}
To better illustrate the interplays that can emerge in real-world devices, we present a vulnerability found in the router D-Link DAP-2310 v1.00\_o772 firmware. We discovered this vulnerability thanks to the protocol-aware fuzzing capabilities of \ourtool{}, which uses targeted mutations driven by the intra- and inter-service dependencies identified in the services running on the firmware.

\begin{figure}[t!]
    \centering
    \includegraphics[width=\columnwidth]{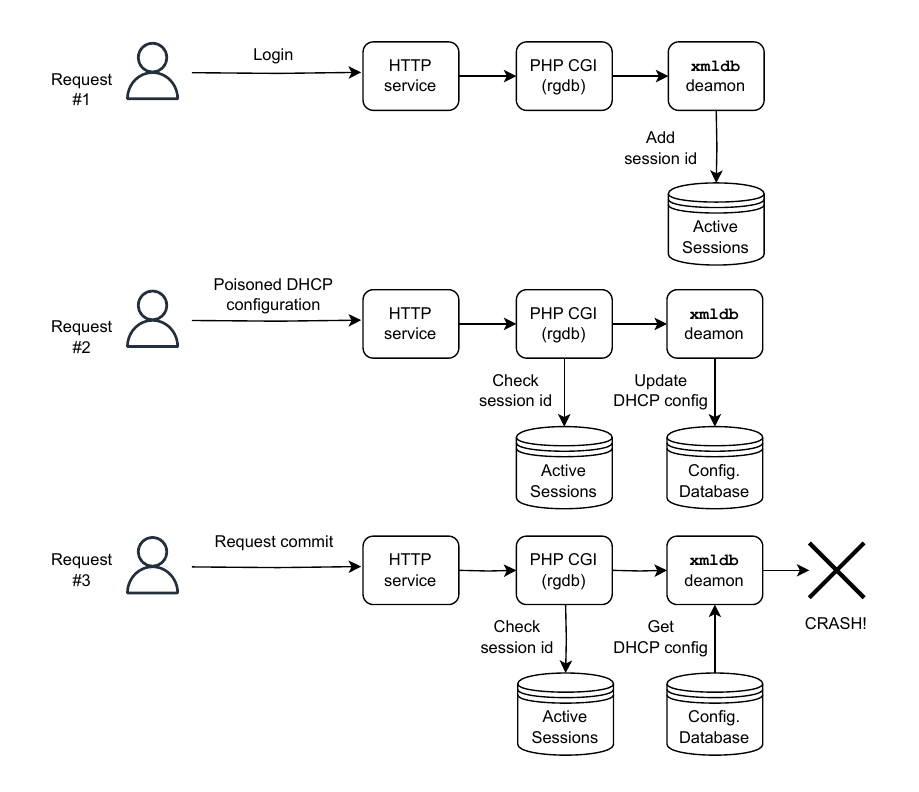}
    \vspace{-2mm}
    \caption{Motivating case study: stack buffer overflow in router D-Link DAP-2310.}
    \label{fig:case-study}
\end{figure}

%
The router runs the lightweight \texttt{httpd} HTTP server\footnote{\texttt{httpd} is \emph{mathopd}, an open-source project available on \href{https://github.com/michielboland/mathopd}{GitHub}, but it is not clear whether the vendor has customized its code and functionalities. The public development from the original author halted in 2024.}, with CGI handlers implemented in PHP that use internal components such as
\texttt{rgdb} (the client-mode interface, typically a symlink to the \texttt{xmldb} binary) and \texttt{xmldb} (a closed-source runtime configuration daemon).
Exploiting the vulnerability requires three coordinated HTTP requests, exemplified in Figure~\ref{fig:case-study},
affecting three
different processes. In more detail:

\begin{enumerate}[label=\protect\circnum{\arabic*}]
  \item {\em Authentication.}     First, login is performed through the HTTP interface, establishing an authenticated session (internal session id) used by the server-side web stack:
    {\small
    \begin{verbatim}
POST /login.php HTTP/1.1
Host: 192.168.0.50
Content-Type: application/x-www-form-urlencoded

ACTION_POST=LOGIN&LOGIN_USER=admin&LOGIN_PASSWD=&
login=Login
    \end{verbatim}
    }\vspace{-4mm}

  \item {\em Poisoned {\tt xmldb} state via DHCP configuration.} Second, a crafted DHCP configuration is submitted via an HTTP POST request. This configuration is passed to a PHP CGI script that, in turn, forwards it to \texttt{xmldb} deamon through the \texttt{rgdb} interface. The \texttt{xmldb} deamon
  records the overlong string into its configuration tree (persisted to disk) without proper validation, inserting the overlong string highlighted in black:  
    \begin{small}
\begin{Verbatim}[commandchars=\\\{\}]
POST /adv_dhcpd.php HTTP/1.1
Host: 192.168.0.50
Content-Type: application/x-www-form-urlencoded

ACTION_POST=adv_dhcpd&srv_enable=1
&ipaddr=192.168.0.30\colorbox{black}{\textcolor{white}{[Insert >1024 bytes]}}
&iprange=235&ipmask=255.255.255.0&...
\end{Verbatim}
    \end{small}

  \item {\em Exploit buffer overflow in \texttt{xmldb} via configuration.}      Finally, the configuration changes can be committed using an HTTP GET to the XGI handler (e.g., \texttt{cfg\_valid.xgi}) with an \texttt{exeshell=submit COMMIT} parameter. The handler writes a \texttt{submit COMMIT} command into the running \texttt{xmldb} daemon's control interface (pipe/socket/FIFO). Then, \texttt{xmldb} loads the poisoned value
  and copy it unsafely, resulting in a stack buffer overflow:
    \begin{small}
\begin{Verbatim}[commandchars=\\\{\}]
GET /cfg_valid.xgi?
    random=...&exeshell=submit%20COMMIT
    &exeshell=submit%20DHCPD&flag=1 HTTP/1.1
Host: 192.168.0.50
\end{Verbatim}
    \end{small}
 The vulnerable code
 performs an unsafe {\tt strcpy}:
    {\small
    \begin{verbatim}
char acStack_410[1024];
...
strcpy(acStack_410, __nptr);
    \end{verbatim}
    }\vspace{-2mm}
\end{enumerate}

This real-world case study clearly shows how an HTTP service may act as the central control plane for device configuration. Yet triggering a serious memory corruption bug requires cross-process stateful interactions involving internal daemons accessed indirectly through HTTP CGI endpoints.

\paragraph{Challenges.}
As seen in the motivating case study, embedded firmware images may not be a single monolithic blob but a complex interplay of cooperating services, daemons, and kernel components sharing mutable state via IPC, filesystems, and runtime databases. Unfortunately, there are several challenges that limit the effectiveness of state-of-the-art fuzzing solutions when considering these complex firmware systems:

\begin{description}
    \item[C1: Stateful and multi-request interactions.]  
    Interactions with modern devices span several requests, such as login, token fetch, sending configuration parameters, and saving the overall configuration. Each of these requests builds on the previous ones, making them inherently stateful. State-of-the-art firmware fuzzers, such as TriforceAFL or FirmAFL, ignore such custom protocols, blindly destroying valid request sequences, thus failing to reach interesting states in most of their fuzzing attempts. Protocol-aware fuzzers, such as AFLNet, can perform significantly better on firmware systems such as the case study but have not been ported to the firmware analysis context.


    \item[C2: Multiple interdependent services.]  
    While a user may interact with a single border service, e.g., the HTTP handler, modern devices are characterized by complex interdependent services. Hence, data sent to a network-facing service may traverse different processes, passing through IPC channels and persistent storage, and then finally reach a vulnerable service. State-of-the-art firmware fuzzers do not currently keep track of how input bytes, received from the network, may impact the overall execution of the entire firmware system. Protocol-aware fuzzers, being mostly designed for single-process applications, cannot cope with such multi-service complex dynamics.


    \item[C3: Execution overhead from system mode emulation.]  
    A major problem in firmware analysis is the execution overhead introduced by system mode emulation. While the previous challenges may apparently already find some solution ingredients in existing works, such as protocol-aware fuzzing, a significant problem is that these prior proposals cannot be blindly ported to the context of firmware analysis due to the huge cost of fuzzing attempts. Hence, a major challenge is how to be protocol-aware and track complex intra- and inter-service dependencies while minimizing the overhead. In particular, existing fuzzing strategies may require a number of attempts (i.e., requests) that is unsustainable when using system mode emulation, thus requiring optimizations and minimization strategies.

\end{description}

\section{Stateful Taint-Assisted Full-system Fuzzing}
\label{sec:approach}

In this section, we present \ourtool{}, a firmware fuzzing framework designed to discover bugs in Linux-based firmware software. We first present the overall architecture of \ourtool{}, contextualizing it within the challenges presented in Section~\ref{sec:motivation}, and then provide detailed coverage of its most important conceptual components.


\subsection{Overall Approach}

\begin{figure}
    \centering
    \includegraphics[width=0.95\columnwidth]{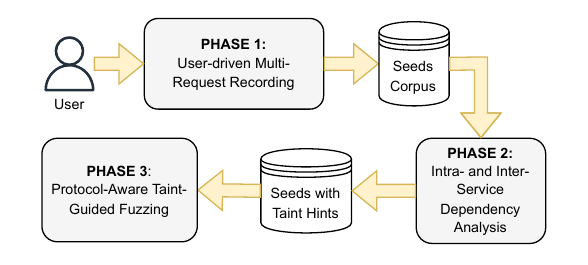}
    \vspace{-4mm}
    \caption{Bird's-eye view of the conceptual phases in \ourtool{}.}
    \label{fig:overall_approach}
\end{figure}

\ourtool{} is structured into three main conceptual phases: {\em User-driven Multi-Request Recording} (Section~\ref{ssec:exploration_corpus_gathering}), {\em Intra- and Inter-Service Dependency Analysis} (Section~\ref{ssec:dataflow_analysis}), and {\em Protocol-Aware Taint-Guided Fuzzing} (Section~\ref{ssec:fuzzing}). Figure~\ref{fig:overall_approach} 
provides a bird's-eye view of these phases, which we now review at a high level.


In the first phase, {\em User-driven Multi-Request Recording}, the user interacts with the firmware under test by manually performing different interactions based on the intended functionalities of the device. As in prior works~\cite{triforceafl, zheng2019firm,  srivastava2019firmfuzz},
in this article, we focus our attention on interactions primarily carried out through the HTTP interface exposed by the firmware. However, different from prior works, users are not forced to restrict interactions to those affecting only the HTTP daemon but are free to explore the functionalities, since \ourtool{} targets testing of {\em all affected programs} in the system. Hence, the ultimate goal of this phase is to exercise as many firmware functionalities as possible, such as configuration endpoints, status interfaces, or management operations, allowing \ourtool{} to capture a representative \textit{seed corpus} of request sequences. Our framework automatically logs these requests at the packet level (e.g., in PCAP format), tags them with timing and response metadata, and stores them in a \textit{seed corpus}. This corpus allows \ourtool{} to replay these recorded sequences in later phases, faithfully reproducing the execution context and ensuring that subsequent fuzzing operates on exactly the same firmware state that the user analyst explored. Prior works have assumed the availability of such a corpus, without systematically providing support for building it. In contrast, \ourtool{} makes this crucial preliminary phase an essential and well-structured component of its architecture.


In the second phase, {\em Intra- and Inter-Service Dependency Analysis}, \ourtool{} replays each recorded sequence and instruments the firmware execution to extract detailed runtime information. During this replay, it tracks basic block coverage, memory accesses, and filesystem activities. In particular, a taint-based dataflow tracker logs how input bytes propagate through CPU registers, memory, IPC channels, and filesystem operations. This analysis reveals which parts of the input requests may influence which parts of the firmware code, highlighting in detail the intra- and inter-service dependencies impacting the firmware execution. To the best of our knowledge, no prior work in firmware fuzzing has explored a similar analysis with the same fine granularity.
The ultimate goal of this phase is to generate a seed corpus annotated with {\em taint hints}, i.e., valuable information about the firmware execution dependencies that could be exploited during fuzzing to reach interesting execution states.



In the final phase, {\em Protocol-Aware Taint-Guided Fuzzing}, \ourtool{} sends mutated request sequences to the firmware, preserving their ordering and context as observed during the user interactions, while monitoring the firmware execution in search for crashes and other indicators of inconsistent running state. Mutations are protocol-aware and are performed exploiting the taint hints collected in the previous phase. Code coverage 
is tracked to evaluate whether non-crashing inputs should be added to the current corpus for further mutations. 
Due to the large overhead introduced by system-mode emulation, \ourtool{} adopts different optimizations, including multi-staged forkservers to efficiently checkpoint complex protocol states. The ultimate goal of this phase is to possibly generate input request sequences able to reach deep execution paths and surface subtle, inter-service, state-dependent vulnerabilities.


When considering the three challenges mentioned in Section~\ref{sec:motivation}, \ourtool{} tackles them in the following way:

\begin{description}
  \item[C1: Stateful and multi-request interactions.] \ourtool{} has been designed to be protocol-aware, capturing full request chains, including authentication, token exchanges, and configuration updates, preserving the exact inter-request context required to trigger deep logic paths. 

  \item[C2: Multiple interdependent services.]\hspace{4pt} \ourtool{} exploits whole system taint analysis to track user‐controlled bytes through all user-space processes,
  IPC channels, shared memory, and filesystem operations. This unified view produces fine‐grained byte‐to‐code mappings and uncovers both volatile and persistent dependencies across service boundaries. 

  \item[C3: Execution overhead from system mode emulation.] \ourtool{} introduces different optimization and minimization strategies to make its analysis scalable and sustainable in the context of system-mode firmware analysis, including {\em approximated inter-region dependency tracking} and {\em multi-staged forkservers}. 
\end{description}

\subsection{User-driven Multi-Request Recording}
\label{ssec:exploration_corpus_gathering}

\begin{figure}[t]
    \centering
    \vspace{-3mm}
    \includegraphics[width=0.95\columnwidth]{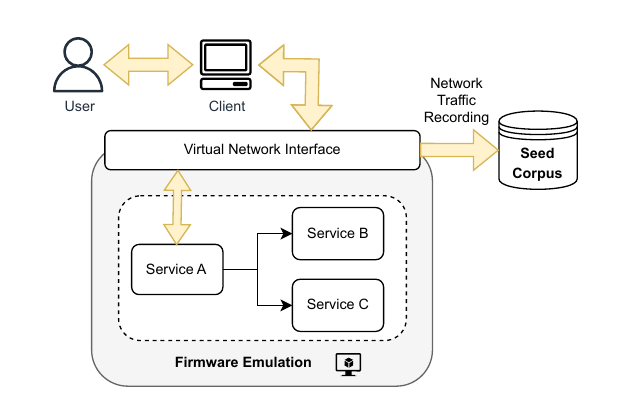}
    \vspace{-3mm}
    \caption{{\em User-driven Multi-Request Recording} Phase: A client communicates with the firmware through a virtual network interface, exercising externally reachable services exposed by the firmware under emulation. \ourtool{} records the network interactions ({\em seed corpus}) to allow replay in the next phases.}
    \label{fig:exploration_phase}
\end{figure}


A crucial requirement for performing effective fuzzing of software is the availability of a rich and diverse seed corpus. Unfortunately, in the context of firmware analysis, it is quite rare to obtain a set of benign input request interactions for testing from the manufacturer. For this reason, as in AFLNet~\cite{pham2020aflnet},
\ourtool{} relies on input request sequences manually generated by a user analyst during an emulation session monitored by \ourtool{}. 
In contrast to prior works, 
\ourtool{} systematizes this acquisition procedure, making it part of the system and thus providing adequate support to the end user. 

Figure~\ref{fig:exploration_phase} graphically depicts the expected workflow during this phase.
Given a firmware image, the emulation layer of \ourtool{} unpacks it, patches or rebuilds kernels when needed, and mounts the reconstructed root filesystem with valid device nodes and init scripts, configuring a system-mode emulation with virtualized CPU, memory, common peripherals, and injected dummy drivers or stubs for hardware components that cannot be emulated precisely (e.g., GPIOs, watchdogs, UARTs). Finally, a virtual network interface is made available to expose firmware services to external clients.

When the emulated environment is fully ready, the user analyst is expected to perform meaningful interactions using a client, e.g., a browser, towards the virtual network interface where \ourtool{} has made available the different network services spawned by the firmware. Our current implementation of \ourtool{} is tailored towards interactions targeting the HTTP service. These interactions are recorded as network packet traces, where each trace reflects a client session captured to preserve protocol structure, ordering, authentication flows, and state transitions. Given the large number of HTTP requests possibly recorded during the interactions, \ourtool{} performs a preliminary filtering of likely uninteresting requests, discarding endpoints related to static files, such as Javascript code, images, and style sheets.

The recorded sessions form the initial \emph{seed corpus} that are replayed in the next phase to identify interesting service interactions.

\paragraph{Example.} 
During an emulation session of the firmware D-Link DAP-2310 v1.00\_o772, the analyst used a web browser to interact with the emulated HTTP service exposed by the virtual network interface. The session began with a \texttt{GET} request to \texttt{/login.php} to retrieve the login page, followed by a \texttt{POST} request submitting the login form with the default \texttt{admin} username and an empty password to authenticate the user. Upon successful login, the analyst navigated through several configuration pages, including \texttt{/index.php} (dashboard overview), \texttt{/home\_sys.php} (system status), and \texttt{/adv\_dhcpd.php} (DHCP server settings), the latter receiving a \texttt{POST} request to modify DHCP server parameters such as IP range, subnet mask, and lease time. The interaction concluded with multiple requests to firmware-specific endpoints such as \texttt{/cfg\_valid.xgi}, \texttt{/cfg\_valid.php}, and \texttt{/\_\_action.php}, which are used to apply and commit configuration changes. All HTTP exchanges were recorded as part of the initial seed corpus, preserving the protocol order, authentication flow, and session state transitions for subsequent fuzzing stages.

\subsection{Intra- and Inter-Service Dependency Analysis}
\label{ssec:dataflow_analysis}

\begin{figure*}[t]
    \centering
    \includegraphics[width=0.9\textwidth]{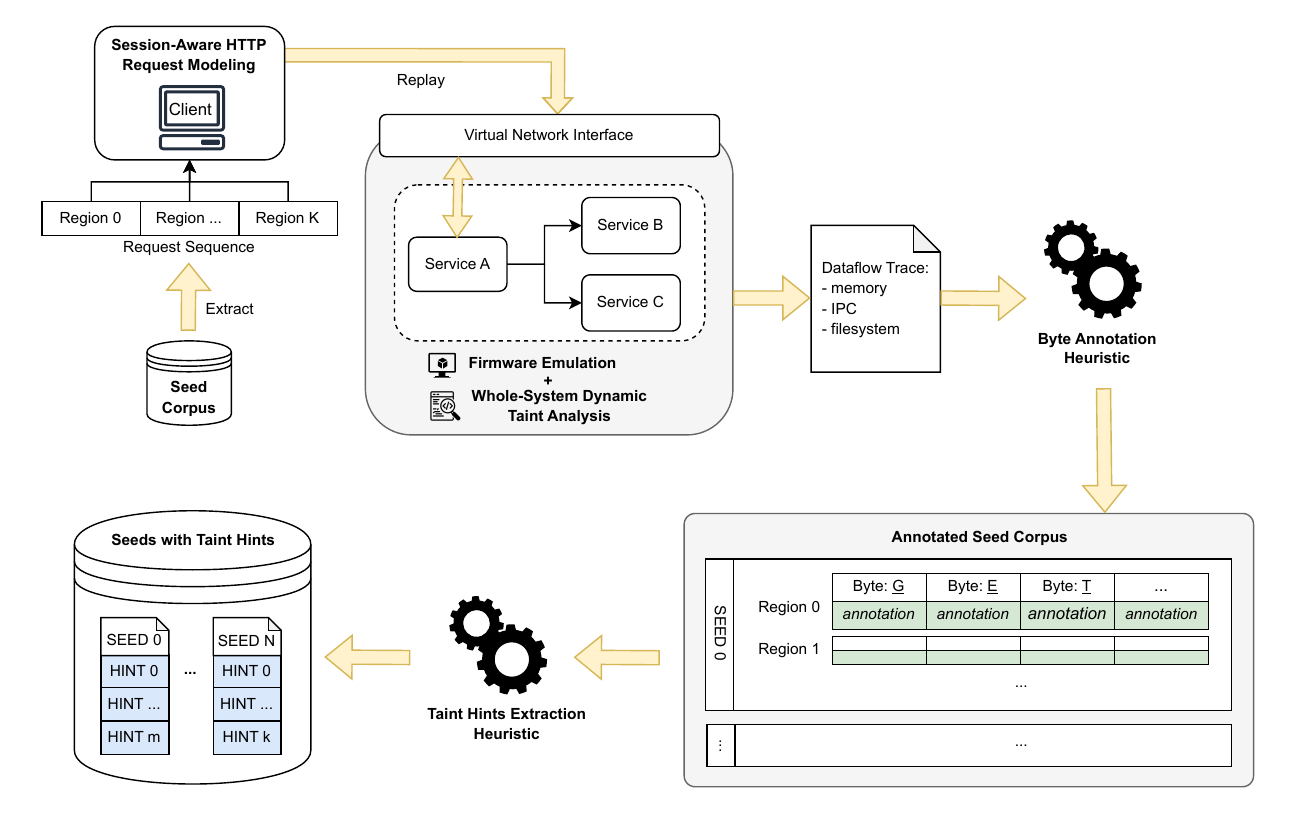}
    \caption{Intra- and Inter-Service Dependency Analysis: (a) network interactions are extracted from the seed corpus; (b) split into regions (requests); (c) replayed using a {\em session-aware HTTP request modeling} component within an emulated environment where {\em whole-system dynamic taint analysis} is performed to track how region bytes flow into memory, IPC channels, files, and more generally impact firmware service execution; (d) the {\em byte annotation heuristic} enriches each byte of each region with annotations summarizing its impact on the firmware execution; (e) the {\em taint hints extraction heuristic} transforms the annotations into actionable {\em taint hints} to guide the mutations during fuzzing.}
    \label{fig:pre_analysis_phase}
\end{figure*}

After collecting a rich corpus of network interactions, \ourtool{} initiates the intra- and inter-service dependency analysis, whose main workflow is depicted in Figure~\ref{fig:pre_analysis_phase}. The ultimate goal of this analysis is to generate actionable and valuable {\em taint hints} that can help the fuzzing engine to perform effective mutations based on the intra- and inter-service dependencies identified during the analysis. In the remainder of this section, we provide details about each step executed during the analysis.

\subsubsection{Session‐Aware HTTP Request Modeling}
\label{sssec:request-modeling}

After extracting a seed, i.e., a network interaction, from the corpus, \ourtool{} splits it into regions, i.e., distinct network requests, and then processes them with the {\em session-aware HTTP request modeling} component to reconstruct a structured view of the HTTP requests. Such processing is performed through a 
HTTP parser that manages a correct \textit{request reassembly} by inferring its termination using HTTP protocol-defined rules:
\begin{itemize}
\item For Content-Length-based requests, byte-precise parsing is used.
\item Chunked requests are reconstructed by parsing each chunk until termination.
\item In the absence of explicit length, connection close semantics are respected.
\item For markup-heavy responses (HTML/XML), logical end-tags (e.g., \texttt{</html>}) are additionally used to infer completeness.
\end{itemize}

Differently from timeout-based strategies used in prior protocol-aware works (see Section~\ref{ssec:protocol-fuzzing}),
such as AFLNet, our approach handles partial, malformed, or delayed responses gracefully by relying on protocol-aware parsing rather than timing heuristics, improving robustness under real-world conditions. This improved robustness makes it possible to significantly improve the success rate of network interaction reproduction during later stages. 
%
%
Moreover, by obtaining a structural view of the HTTP requests, \ourtool{} can focus its dependency analysis and fuzzing mutations on the significant parts of the network interactions.

\paragraph{Example.} 
During the replay of the previously recorded interaction with the D-Link DAP-2310 v1.00\_o772 firmware, the request \texttt{GET /index.php} was issued to the emulated HTTP web server. The server responded with a long, chunked HTTP response containing the HTML structure and embedded JavaScript code of the device’s administration dashboard. The payload was split across multiple TCP packets, starting with the \texttt{Transfer-Encoding:~chunked} header and spanning over several kilobytes until the closing \texttt{</html>} tag.
In timeout-based request reassembly strategies, as adopted by AFLNet, chunked responses may be truncated prematurely. This results in incomplete reassembly of the intended response. The remaining chunks can then be mistakenly associated with the following request, as the first data received after issuing it may still belong to the unfinished previous response. In contrast, \ourtool{}’s session-aware HTTP request modeling correctly reassembles the response by parsing the HTTP headers, recognizing the chunked transfer encoding, and reading each chunk until completion, with an additional safeguard of detecting logical HTML end-tags.

\subsubsection{Whole System Taint Analysis}
\label{ssec:streaming}

While parsing the network interactions through the session-aware HTTP request modeling component, \ourtool{} replays the identified structured regions (requests) into an emulated environment running the firmware under test. During the emulation, \ourtool{} performs whole-system dynamic taint analysis aimed at tracking how region bytes propagate through memory, IPC channels, and filesystem operations. 

One major design choice in \ourtool{} is how many distinct taint sources should be tracked. Ideally, we would like to track how each input (region) byte propagates throughout the execution; however, this approach does not scale: binary-level dynamic taint analysis typically supports only a limited number of taint sources to limit the time and space overhead, 
e.g., up to 8 distinct taint sources. With a limited number of taint sources, a strategy could be to perform several executions, selectively tainting only a few bytes per execution. However, such an approach does not scale since it requires repeating the same execution a large number of times (i.e., $N / n$ executions, where $N$ is the total number of bytes across all regions and $n$ is the number of distinct taint sources) while considering different subsets of tainted bytes from the input. For these reasons, \ourtool{} adopts a more scalable approach where a single taint source is used, exploiting a byte annotation heuristic (Section~\ref{sssec:byte-annotation}), inspired by the input-to-state relationships from REDQUEEN (Section~\ref{sec:background}), to refine the taint dependencies.


During taint analysis, \ourtool{} generates two types of events: {\em tainted memory accesses} and {\em filesystem-based dependencies across regions}. For both types of events, \ourtool{} exploits the concept of a {\em processing time window} of a region, which starts at the time when the HTTP request related to the region is received by the firmware and ends at the time of receipt of the first subsequent HTTP response.

\paragraph{Memory accesses.}
During execution, \ourtool{} emits byte-granular tainted memory access events,
where each event captures the following information:
\begin{itemize}
    \item \texttt{region\_id}: identifier of the region whose processing time window contains the memory access.

    \item \texttt{pc}: 
    identifier combining the inode\footnote{The PC of the basic block may not be unique across different processes, hence, \ourtool{} uses the filesystem inode number of the application binary to discriminate the same PC across different applications.} and the program counter (PC) of the basic block performing the memory access.
    \item \texttt{gpa}: the 
    physical address of the memory access.
    \item \texttt{value}: the tainted (loaded or stored) byte value.
\end{itemize}

Each event is stored in a per-seed data structure {\tt tainted\_} {\tt memory\_events}: {\tt tainted\_memory\_events[load]} collects the {\em load} events, while {\tt tainted\_memory\_events[store]} collects the {\em store} events.

%
%
%
%

\paragraph{IPC channels.}
\ourtool{} can naturally track tainted bytes traversing inter-process communication (IPC) channels thanks to its taint analysis being able to observe memory regions (even the ones used by the kernel to store temporary data) or other virtual files involved in IPC channels.

\paragraph{Filesystem operations.} \ourtool{} tracks which filesystem operations are executed by the firmware while processing a specific region and then establishes reader-writer relationships whenever a (reader) region accesses a file previously written or created by a (writer) region.
In more detail, syscalls are split into two semantic classes:
\begin{itemize}
  \item {\em Writer syscalls:} operations that create or modify file content or metadata, such as \texttt{write}, \texttt{open} with write flags, \texttt{rename}, \texttt{unlink}, or \texttt{truncate}.
  \item {\em Reader syscalls:} operations that access existing file content or metadata, such as \texttt{read}, \texttt{open} with read-only flags, \texttt{stat}, \texttt{lstat}, or \texttt{fstat}.
\end{itemize}
Given these two semantic classes, \ourtool{} identifies a dependency between two regions $R_i$ and $R_j$ within the same seed when during the processing
of a region $R_i$ a writer syscall is executed to a file node $f_k$,
and, later, during the processing time window of region $R_j$ a reader syscall is executed on the same file node $f_k$. More formally, the inference rule is:
\[
R_i \xrightarrow{\texttt{writer}} f_k \xrightarrow{\texttt{reader}} R_j \quad\Rightarrow\quad R_i \xrightarrow{f_k} R_j
\]
\ourtool{} stores these types of region dependencies 
within a per-seed map {\tt region\_deps}, where each reader-writer pair ($R_j$, $R_i$) is mapped to the list of involved files:
\[
\texttt{region\_deps}[R_j][R_i] = [\ldots, f_k, \ldots]
\]

\begin{figure*}[t]
    \centering
    \includegraphics[width=\textwidth]{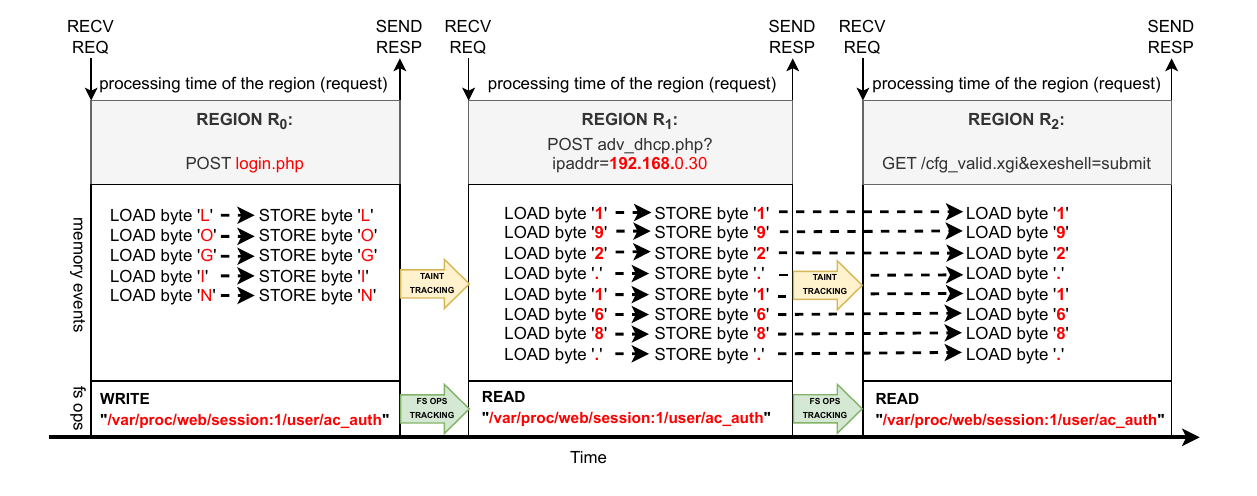}\vspace{-5mm}
    \caption{Example of whole system taint analysis: the file {\tt /var/proc/web/session:1/user/ac\_auth} is created (writer) by the first region (first request) and then read (readers) by the two subsequent regions (second and third requests); the bytes {\tt 192.168.} from the second region are propagated to the third region.}
    \label{fig:example-system-taint}
\end{figure*}

\paragraph{Example.} Figure~\ref{fig:example-system-taint} depicts an example of memory and file system tracking performed by \ourtool{} during whole system taint analysis. In particular, the example considers an interaction constituted by three regions:
\begin{enumerate}
  \item[$R_0$] When processing the request {\tt POST login.php}, the program first loads the bytes {\tt LOGIN} from the request body of $R_0$ and then stores them into a different memory area. \ourtool{} collects several memory events to capture these loads and stores of tainted bytes from the region. Meanwhile, the program also creates the file {\tt /var/proc/web/session:1/user/ac\_auth}, thus acting as a writer, an operation tracked by \ourtool{}. Note that the window processing time of a region starts with the receipt of the related request and ends with the first response.
  \item[$R_1$] When processing the request {\tt POST adv\_dhcp.php?ipaddr}\break{\tt =192.168.0.30}, \ourtool{} tracks how the tainted bytes {\tt 192.168.} from region $R_1$ are moved around in memory. Also, \ourtool{} marks the second region $R_1$ as a reader of the file $f$ with path {\tt /var/proc/web/session:1}\break{\tt /user/ac\_auth} previously written by region $R_0$, i.e., $R_0 \xrightarrow{f} R_1$.
  \item[$R_2$] When processing the request {\tt GET /cfg\_valid.xgi\&}\break{\tt exeshell=submit}, \ourtool{} identifies how some bytes ({\tt 192.168.}) from region $R_1$ are loaded during the processing of region $R_2$. Finally, it also marks the third region $R_2$ as a reader of the file $f$ with path {\tt /var/proc/web/session:1/user/ac\_auth} previously written by region $R_0$, i.e., $R_0 \xrightarrow{f} R_2$.
\end{enumerate}

\subsubsection{Byte Annotation Heuristic}


\label{sssec:byte-annotation}

\begin{algorithm}[!ht]
  \SetInd{0.7em}{0.7em} 
  \SetAlgoNoLine
  \SetAlgoNoEnd 
  \DontPrintSemicolon

  \caption{Byte Annotation Heuristic}\label{alg:byte-annotation-emilio}
  
  \SetKwInOut{Input}{Input}
  \SetKwInOut{Output}{Output}
  
  \Input{Seeds $S$, where $\forall s \in S$:\\- $s$.{\tt regions} (seed bytes for each region)\\- $s$.{\tt tainted\_memory\_events} (list)\\- $s$.{\tt region\_deps} (per region map)} 
  \Output{Annotated seeds $S$, where $\forall s \in S$:\\- $s$.{\tt regions} (seed bytes for each region)\\- $s$.{\tt tainted\_memory\_events} (list)\\- $s$.{\tt region\_deps} (per region map)\\- $s$.{\tt annotated\_bytes} (annotated bytes)}
  \BlankLine
  $PCs \leftarrow$ \{\} \tcp*{mapping: pc $\mapsto$ seed} \label{line:init-pcs}
  \For{${s} \in {S}$}{ \label{line:loop-per-seed}
    FilterAndFlattenRegionDeps($s$)\; \label{line:filter-region-first}
    $T \leftarrow$ IndexRegionByteSequencesInTrie($s$)\; \label{line:index-bytes-sequences}

    

    \For{${r} \leftarrow 0$ \KwTo $\mathrm{len}({s}.\texttt{regions})-1$}{ \label{line:init-ann-start}
      $s$.{\tt annotated\_bytes}[$r$] $\leftarrow$ \\ 
      $~~$ [\{{\tt deps}: \{\}, {\tt PCs}: \{\}\} : $b \in s.\texttt{regions}[r]$]\;
    } \label{line:init-ann-end}
    

    \For{${t} \in$ \{load, store\}}{ \label{line:for-type-start}
    $E \leftarrow$ []\; \label{line:init-E}
    \ForEach{${e} \in {s}.\texttt{tainted\_memory\_events}[t]$}{\label{line:stream-events}
          ProcessMemEvent($s, e, E, T, PCs$)\; \label{line:process-event-invocation}
        }
    } \label{line:for-type-end}
  }

  
  \Return $S$\;
\end{algorithm}



After collecting events during whole-system taint analysis, \ourtool{} exploits the \textit{Byte Annotation Heuristic} (Algorithm~\ref{alg:byte-annotation-emilio}) to enrich each region's bytes with metadata describing how bytes have propagated through memory, which basic blocks accessed them, and which inter-region dependencies related to filesystem operations exist. As mentioned in Section~\ref{ssec:streaming}, since \ourtool{} adopts a dynamic taint analysis with a single taint label to make the approach more scalable, the heuristic has to infer how specific byte sequences impact specific basic blocks using a value-matching strategy, inspired by the input-to-state relationships from REDQUEEN (Section~\ref{sec:background}), to refine the taint dependencies. 

The heuristic takes as input a set of seeds $S$ that must be annotated. Each seed $s$ is composed of:
\begin{itemize}
    \item \texttt{regions}: arrays of bytes, one for each input region;
    \item \texttt{tainted\_memory\_events}: a chronologically ordered list of tainted memory access events (Section~\ref{ssec:streaming});
    \item \texttt{region\_deps}: a per-seed data structure storing filesystem-related region dependencies (Section~\ref{ssec:streaming}).
\end{itemize}
The output is the same set of seeds S, where each seed has been enriched with:
\begin{itemize}
    \item \texttt{annotated\_bytes}: for each byte in each region, a structure containing the set of identifiers ({\tt deps}) for the regions that access the byte during processing and the set of program counters ({\tt PCs}) related to the basic blocks that access it.
\end{itemize}


Initially, the heuristic initializes the map \texttt{PCs} (line~\ref{line:init-pcs}). This map tracks, for each basic block (identified by its program counter), which seed first\footnote{The seed ordering does not matter.} executed that block and generated a byte annotation.
The algorithm then considers each seed $s$ from $S$ (line~\ref{line:loop-per-seed}) and performs three main steps:
\begin{enumerate}
    \item {\it Filtering and Flattening Inter-Region Dependencies.} The algorithm filters the raw region dependencies $s$.\texttt{region\_deps} by invoking the procedure \textit{FilterAndFlattenRegionDeps()} at line~\ref{line:filter-region-first} (Algorithm~\ref{alg:filter-region-deps}). The goal of this filtering procedure is to retain only the most relevant region dependencies by eliminating redundant or less significant associations.

    \item {\it Multi-Region Subsequence Indexing.} The algorithm builds a per-seed trie\footnote{A trie~\cite{fredkin1960trie} is a specialized search tree where each node represents a character and stores indices indicating where the string from root to node appears.} $T$ at line~\ref{line:index-bytes-sequences} to track the positions of all byte subsequences from the seed's regions using the procedure {\em IndexRegionByteSequencesInTrie}. The trie $T$ is used in the next step during value-matching to determine whether a given byte sequence appears in any region. 


    \item {\it Region Byte Annotation.} The algorithm initializes the annotations $s$.\texttt{annotated\_bytes} for each byte of each region at lines~\ref{line:init-ann-start}--\ref{line:init-ann-end}. It then processes the memory events in $s$.\texttt{tainted\_memory\_events} at lines~\ref{line:for-type-start}--\ref{line:for-type-end}, distinguishing between {\em load} and {\em store} events, to generate the annotations using the procedure \textit{ProcessMemEvent} (Algorithm~\ref{alg:process-mem-events}).
\end{enumerate}
We now review in more detail the procedures used by these three steps, explaining their goals and motivation.
%

\smallskip

\begin{algorithm}[t!]
  \SetInd{0.7em}{0.7em} 
  \SetAlgoNoLine
  \SetAlgoNoEnd 
  \DontPrintSemicolon

  \caption{Filter and Flatten Region Deps}\label{alg:filter-region-deps}
  \SetKwProg{Proc}{Procedure}{}{}
  \Proc{{\rm FilterAndFlattenRegionDeps($s$)}}{
      \textit{deps} $\gets s$.\texttt{region\_deps}\;
      \textit{MWF}$[f] \gets\min\{w\mid\exists\,r: f\in {deps}[r][w]\}$\; \label{line:init-mwf-f}
      \textit{FW}$[w]\gets \{f \mid $\textit{MWF}$[f] = w\}$\;  \label{line:init-fw-w}
      $W[r] \gets \{w \mid w \in {deps}[r]\}$\; \label{line:init-w-r}
      \textit{deps}$[r]\gets\{$BuildDeps($r$,\textit{W},\textit{FW},$\{\})\,|\,r\in $ \textit{deps}\}\;
      \label{line:foreach-r-deps-end}
  }
\end{algorithm}

\begin{algorithm}[t!]
  \SetInd{0.7em}{0.7em} 
  \SetAlgoNoLine
  \SetAlgoNoEnd
  \DontPrintSemicolon

\caption{Build Filtered Dependencies}\label{alg:build-filtered-deps-map}
\SetKwProg{Proc}{Procedure}{}{}
\Proc{{\rm BuildDeps($r$, \textit{W}, \textit{FW}, $V$)}}{ 
    {\bf if} $r \in V$ {\bf then return} $\{\}$ {\bf else} $V$.insert($r$)\;\label{line:visited-check}
    \textit{deps} $\gets$ \{\} \tcp*{mapping: writer $\mapsto$ set(files)}
    \ForEach{$w \in {\rm sorted}(W[r])$}{ \label{line:foreach-w}
        \If{$w \in $ \textit{FW}}{ \label{line:attach-fw-start}
            \textit{deps}[$w$].\text{add}($f$) \quad $\forall f \in \textit{FW}[w]$\; \label{line:attach-fw-end}
        }
        \textit{wdeps} $\leftarrow$ BuildDeps($w$, \textit{W}, \textit{FW}, $V$)\; \label{line:recursive-call}
        \ForEach{ww $\in$ wdeps}{ \label{line:merge-child-start}
            \textit{deps}[$ww$] $\gets$ \textit{deps}[$ww$] $\cup$ \textit{wdeps}[$ww$]\;
        } \label{line:merge-child-end}
    }
    \Return \textit{deps}\; \label{line:return-filtered-wdeps}
}
\end{algorithm}

\begin{figure*}[!t]
    \centering
    \includegraphics[width=0.8\textwidth]{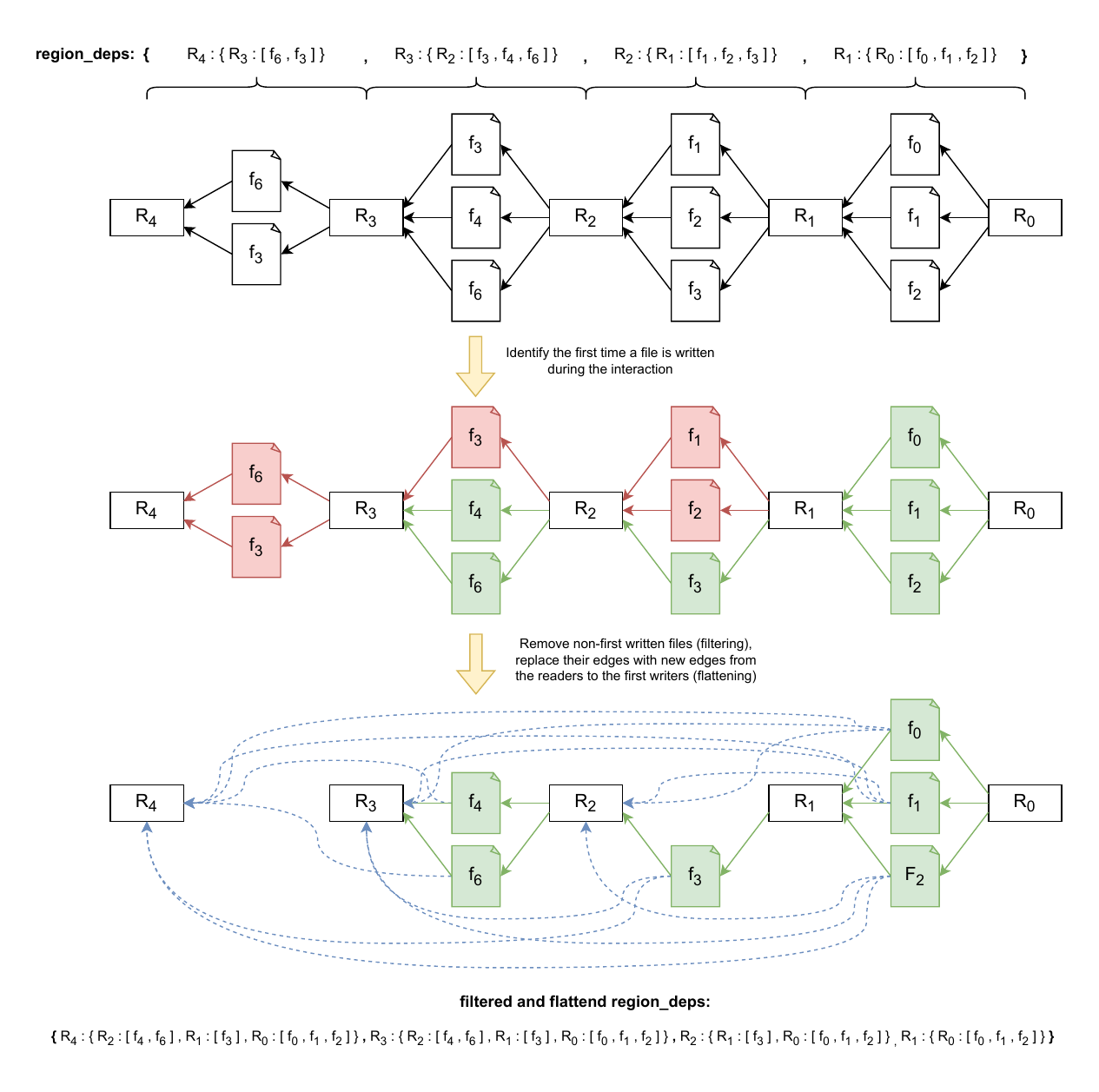}
    \caption{Example of the filtering and flattening of inter-region dependencies ({\tt region\_deps} from each seed) performed by the procedure {\em FilterAndFlattenRegionDeps()} (Algorithm~\ref{alg:filter-region-deps}).\label{fig:filter-regions-deps}}
    \label{fig:placeholder}
\end{figure*}

The procedure \textit{FilterAndFlattenRegionDeps()} (Algorithm~\ref{alg:filter-region-deps}) filters and flattens the inter-region dependencies \texttt{region\_deps} from each seed. Figure~\ref{fig:filter-regions-deps} provides a high-level intuition of the work performed by the procedure\footnote{We remark that \texttt{region\_deps} maps readers to writers, hence representing writer-reader edges across regions in the opposite direction.}: first, the algorithm identifies the first time files are written during the interaction (green files and edges); then, edges and files related to non-first writers (red files and edges) are removed and replaced with edges connecting the affected readers to the first writers (blue edges). In more detail, \textit{FilterAndFlattenRegionDeps()} operates in four distinct phases.
First, each file $f$ is assigned to the first ({\em minimum}) region writer $\textit{MWF}[f]$, i.e., the smallest region identifier\footnote{Region identifiers reflect the request ordering from an interaction.} that writes $f$ (line~\ref{line:init-mwf-f}). This assignment ensures that each file contributes exactly once to the filtered dependency structure. 
Second, files are grouped by their minimum writer into the $FW$ mapping (line~\ref{line:init-fw-w}). 
Third, a writer mapping $W$ is constructed by inverting the original {\tt region\_deps} edges (line~\ref{line:init-w-r}). 
Finally, for every reader region $r$, the algorithm invokes \texttt{BuildDeps} to compute a filtered, flat mapping from writers to files reachable from that reader (line~\ref{line:foreach-r-deps-end}); this result replaces the original $\textit{deps}[r]$ entry.
The recursive procedure \textit{BuildDeps()} (Algorithm~\ref{alg:build-filtered-deps-map}) returns, for a given reader $r$, a mapping \textit{deps} between its writers and their filtered written files. For each writer $w$ associated with the current reader, the procedure 
\begin{enumerate*}[label=(\roman*)] 
    \item adds the files for which $w$ is the minimum writer by incorporating $\textit{FW}[w]$ into $\textit{deps}[w]$ (lines~\ref{line:attach-fw-start}--\ref{line:attach-fw-end}), 
    \item recursively computes the writer's dependencies \textit{wdeps} (line~\ref{line:recursive-call}), and
    \item merges (via set union) the writer's mapping into the reader's $\textit{deps}$ (lines~\ref{line:merge-child-start}--\ref{line:merge-child-end}).
\end{enumerate*}
Through this process, the procedure eliminates redundant contributions from files referenced by multiple writers. The computed $\textit{deps}$ mapping is returned to the caller (line~\ref{line:return-filtered-wdeps}).

\smallskip

The procedure \textit{IndexRegionByteSequencesInTrie()} implements a standard trie construction algorithm that tracks the positions of all byte subsequences from the seed's regions. We omit the implementation details for brevity, but discuss how to avoid memory explosion during construction in Section~\ref{sec:implementation}.

\smallskip

\begin{algorithm}[t!]
  \SetInd{0.7em}{0.7em} 
  \SetAlgoNoLine
  \SetAlgoNoEnd 
  \DontPrintSemicolon

  \caption{Process Memory Access Events}\label{alg:process-mem-events}
  \SetKwProg{Proc}{Procedure}{}{}
  \Proc{{\rm ProcessMemEvent($s$, $e$, $E$, $T$, \textit{PCs})}}{
    $e\_{prev} \leftarrow E[e.\texttt{type}]$.last()\; \label{line:last-e}
  \If{$e\_{prev} =~${\tt None} {\rm \bf or}
      $(e.\texttt{region\_id} = e\_{prev}\texttt{.region\_id}$
      {\rm \bf and} $e.\texttt{gpa} = e\_{prev}\texttt{.gpa} + 1)$}{ \label{line:check-contig}
        $E[t]$.append($e$)\; \label{line:append-event}
  }
  \Else{
    $r \leftarrow e\_{prev}\texttt{.region\_id}$\; \label{line:set-region}
    AnnotateSubregion($s, E, r, T,$ \textit{PCs})\; \label{line:annotate-call} \label{line:do-annotation}
    ${E} \gets [e]$\; \label{line:e-clear}
  }
  }
\end{algorithm}

\begin{algorithm}[t!]
  \SetInd{0.7em}{0.7em} 
  \SetAlgoNoLine
  \SetAlgoNoEnd 
  \DontPrintSemicolon

  \small\sloppy
  \caption{Annotation of a Subregion Sequence}\label{alg:annotate-subregion-emilio}
  \SetKwProg{Proc}{Procedure}{}{}
  \Proc{{\rm AnnotateSubregion($s$, $E$, $r$, $T$, $\mathit{PCs}$)}}{

        \For{$\ell \leftarrow {\rm len}(E)$ \KwTo 1}{ \label{line:for-len}
          \For{$k \leftarrow 0$ \KwTo $len(E)-\ell$}{ \label{line:for-k}
            \textit{seq} $\leftarrow \left[ E[i].\texttt{value} \;\middle|\; i \in [k : k+\ell] \right]$\; \label{line:extract-subseq}
            \textit{matches} $\leftarrow$ FindByteSeqInRegions($T$, \textit{seq})\; \label{line:trie-find}
            {\bf if} len(\textit{matches}) $\neq$ $1$ \textbf{then continue}\;\label{line:check-match}
            $r^\prime \leftarrow$ \textit{matches}[0].\texttt{region\_id}\; \label{line:match-region}
            $p \leftarrow$ \textit{matches}[0].\texttt{position}\; \label{line:match-pos}
            {\bf if} $r^\prime > r$ \textbf{then continue}\;\label{line:skip-future-region}
            \For{$j \leftarrow 0$ \KwTo $\ell - 1$}{ \label{line:update-dep-start}
              $b \leftarrow s$.{\tt annotated\_bytes}$[r^\prime][p+j]$\;
              $b$.{\tt deps}.insert($r$)\; \label{line:byte-dep}

              $pc \gets E[j]$\texttt{.pc}\;
              {\bf if} $pc \notin$ \textit{PCs} {\bf then} \textit{PCs}$[pc] \gets s$ \label{line:first-seed-for-pc}\;
              {\bf if} \textit{PCs}$[pc] = s$ {\bf then} $b$.{\tt PCs}.insert($pc$)\; \label{line:ann-pc-if-first-seed}
            }  \label{line:update-dep-end}
          }
        }
  }
\end{algorithm}

The procedure {\it ProcessMemEvent()} (Algorithm~\ref{alg:process-mem-events}) analyzes tainted memory access events by grouping contiguous memory accesses into sequences and triggering annotation when sequences are complete. When a new event $e$ arrives, the procedure checks (lines~\ref{line:last-e}--\ref{line:check-contig}) if it continues a contiguous sequence from a previous event of the same type (load or store) by comparing region identifiers and guest physical addresses. If this is the case, it adds the event to the current sequence (line~\ref{line:append-event}). Otherwise, if the event breaks contiguity, the procedure calls {\it AnnotateSubregion()} (line~\ref{line:do-annotation}) to process the completed sequence\footnote{The procedure is also called when processing the last event from s.$\texttt{tainted\_memory\_events}[t]$. The code is omitted for clarity.}, then clears the event buffer $E$ (line~\ref{line:e-clear}) to start a new sequence.
The {\it AnnotateSubregion()} procedure (Algorithm~\ref{alg:annotate-subregion-emilio}) performs the core annotation logic. It extracts byte sequences of decreasing lengths from the completed event sequence $E$ (line~\ref{line:extract-subseq}) and uses the trie $T$ to find matching patterns in the seed regions (line~\ref{line:trie-find}). 
When a unique match is found in an earlier region (ensuring temporal correctness) via lines~\ref{line:check-match}--\ref{line:skip-future-region}, the procedure annotates the corresponding bytes with dependency information (line~\ref{line:byte-dep}). 
Additionally, it conditionally records the program counters of basic blocks containing memory accesses that contributed to the match (line~\ref{line:ann-pc-if-first-seed}), but only when the current seed is the first to execute those basic blocks and generate annotations for them (line~\ref{line:first-seed-for-pc}). This conditional check helps determine which seed should serve as the reference when targeting specific basic blocks.

\subsubsection{Taint Hints Extraction Heuristic}
\label{sssec:taint-hints-heuristic}

\begin{algorithm}[!t]
\SetInd{0.7em}{0.7em} 
\SetAlgoNoLine
\SetAlgoNoEnd
\DontPrintSemicolon

\caption{Taint Hints Extraction Heuristic}\label{alg:taint-hints}
\SetKwInOut{Input}{Input}
\SetKwInOut{Output}{Output}

\Input{Annotated seeds $S$}


\Output{Hints $H[s]$ for each seed $s \in S$}

$H \gets \{s \mapsto [] : s \in S\}$\;\label{line:init-H}
\For{$s \in S$}{ \label{line:begin-for-s}


    $W \leftarrow \{\}$ \tcp*{region writers}\label{line:W-begin}
    \For{$r \gets 0$ \KwTo $|s.\texttt{annotated\_bytes}|-1$}{
        \For{$j \gets 0$ \KwTo $|B|-1$}{
            \For{$w \in B[j]$.{\tt deps}}{
                $W[w] \leftarrow r$\;
            }
        }
    }\label{line:W-end}

    \For{$r \gets 0$ \KwTo $|s.\texttt{annotated\_bytes}|-1$}{ \label{line:for-r}
        
        $D \gets s$.\texttt{region\_deps}[$r$] $\cup$ W[r]\; \label{line:region-deps}
        $P \gets \{\}$ \tcp*{pending data to process for PCs} \label{line:pending-data}
        $B \gets s$.\texttt{annotated\_bytes[$r$]}\; \label{line:annotated-bytes}
        
        \For{$j \gets 0$ \KwTo $|B|-1$}{ \label{line:for-region-index-begin}
            
            \For{$pc \in B[j]$.{\tt PCs}}{ \label{line:for-annotation-pc}

              \If{$pc \not\in P$}{ \label{line:new-pc-if}
                $P[pc] \gets$ \{ {\tt offsets}: [], {\tt deps}: $D$\}\; \label{line:new-pc-then}
              }

              $P[pc]$.\texttt{offsets}.append($j$)\; \label{line:taint-hint-offsets}           
              $P[pc].$\texttt{deps} $\gets P[pc]$.{\tt deps} $\cup B[j]$.{\tt deps}\; \label{line:taint-hint-deps}

              \If{$j = |B|-1$ {\bf or} $pc \not\in B[j+1]$.\texttt{PCs}}{ \label{line:finalize-hint-begin}
                $H[s]$.append(\{\ \\
                  \quad {\tt region} : $r$\; \label{line:set-region-id}
                  \quad {\tt deps} : $P[pc]$.{\tt deps}\;
                  \quad {\tt offset} : $P[pc]$.{\tt offsets}$[0]$\; \label{line:taint-offset}
                  \quad {\tt len} : |$P[pc].${\tt offsets}|\; \label{line:taint-len}
                \})\;
                $P$.remove($pc$)
              } \label{line:finalize-hint-end}

            }
        }
        $H[s] \gets$ SortFilterHints($H[s]$)\; \label{line:sort-filter}
    } \label{line:for-region-index-end}
} \label{line:end-for-s}
\end{algorithm}

\begin{algorithm}[!t]
  \SetInd{0.7em}{0.7em} 
  \SetAlgoNoLine
  \SetAlgoNoEnd
  \DontPrintSemicolon

  \caption{Sort and Filter Hints}\label{alg:filter-hints}
  \SetKwProg{Proc}{Procedure}{}{}
  \Proc{\rm SortFilterHints($H$)}{
    $I \leftarrow \emptyset$\; \label{line:init-covered}
    $\overline{H} \leftarrow []$\; \label{line:init-H-filtered}
    \For{$h \in$ {\rm SortAscByLen}$(H)$}{ \label{line:for-sorting}
        \textit{start} $\leftarrow h$.{\tt offset}\; \label{line:interval-start}
        \textit{end} $\leftarrow h$.{\tt offset} $+ h$.{\tt len} $- 1$\; \label{line:end}
        
        \If{{\rm NoIntervalOverlap}$($\textit{start}, \textit{end}, $I)$}{ \label{line:if-no-overlap}
            $\overline{{H}}$.append($h$)\; \label{line:append-filtered}
            $I$.append([\textit{start}, \textit{end}])\; \label{line:update-covered}
        }\label{line:interval-end}
    }

    \Return $\overline{{H}}$\;
  }
\end{algorithm}

After generating the byte annotations, \ourtool{} computes the {\em taint hints} that will guide the mutations during the fuzzing phase. Each taint hint is associated with a specific seed $s$ and is characterized by the following fields:
\begin{itemize}
  \item {\tt region}: the identifier of the region in $s$ to be mutated.
  \item {\tt offset}: the byte offset within the region where the mutation should begin.
  \item {\tt len}: the number of consecutive bytes starting from \texttt{offset} for which the mutation should be applied.
  \item {\tt deps}: the set of regions that must be preserved during the replay of $s$ since they either influence the mutated region or are influenced by it.
\end{itemize}
Each taint hint represents a sequence of bytes from a region that can impact the execution of a specific basic block and thus could be worth mutating.

The algorithm for taint hints extraction (Algorithm~\ref{alg:taint-hints}) starts by initializing the hints $H$ for all seeds at line~\ref{line:init-H} and then considers each seed $s$ from the annotated seeds $S$ (Section~\ref{sssec:byte-annotation}) at lines~\ref{line:begin-for-s}-\ref{line:end-for-s}. First, it groups the readers by writer in W at lines~\ref{line:W-begin}-\ref{line:W-end}. Then, for each region of the seed, the algorithm tracks some pending data about the influence of the region on a specific basic block using a map $P$ (line~\ref{line:pending-data}), which accumulates data until taint hints are finalized during the processing of all byte annotations (lines~\ref{line:for-region-index-begin}-\ref{line:for-region-index-end}).

During the processing of each byte annotation, the algorithm iterates over the basic blocks {\tt PCs} influenced by the byte (line~\ref{line:for-annotation-pc}). If a basic block $pc$ is not in $P$ (line~\ref{line:new-pc-if}), i.e., the current byte is the start offset of a new taint hint for a basic block, then $P[pc]$ is initialized (line~\ref{line:new-pc-then}), considering the region dependencies $D$ as its initial dependencies $P[pc]$.{\tt deps}. Then, $P[pc]$ is updated by adding the current byte index to the list of {\tt offsets} (line~\ref{line:taint-hint-offsets}) and merging the byte dependencies with $P[pc]$.{\tt deps} (line~\ref{line:taint-hint-deps}). Afterward, the algorithm evaluates whether the pending data for the current $pc$ can be turned into a taint hint at line~\ref{line:finalize-hint-begin} by checking whether the next annotation byte does not influence $pc$, making this byte the last in the pending sequence influencing $pc$, or whether this is the last byte from the region. The new finalized taint hint picks the first offset from $P[pc]$.{\tt offsets} as the starting {\tt offset} (line~\ref{line:taint-offset}) and a {\tt len} that is equal to the number of consecutive bytes influencing $pc$ (line~\ref{line:taint-len}). After adding the taint hint to $H[s]$, the pending data $P[pc]$ for the related $pc$ is cleared.

The last step of the algorithm when processing a region is to sort and filter the hints $H[s]$, which is done through the procedure {\em SortFilterHints()} (Algorithm~\ref{alg:filter-hints}) that sorts the hints by their length (line~\ref{line:for-sorting}), thus favoring shorter byte sequences, and then keeps only non-overlapping hints (lines~\ref{line:interval-start}-\ref{line:interval-end}) when considering their interval ($s$.{\tt offset}, $s$.{\tt offset}+$s$.{\tt len}-1).

\subsection{Protocol-Aware Taint-Guided Fuzzing}
\label{ssec:fuzzing}

\begin{figure*}[t]
    \centering
    \includegraphics[width=\textwidth]{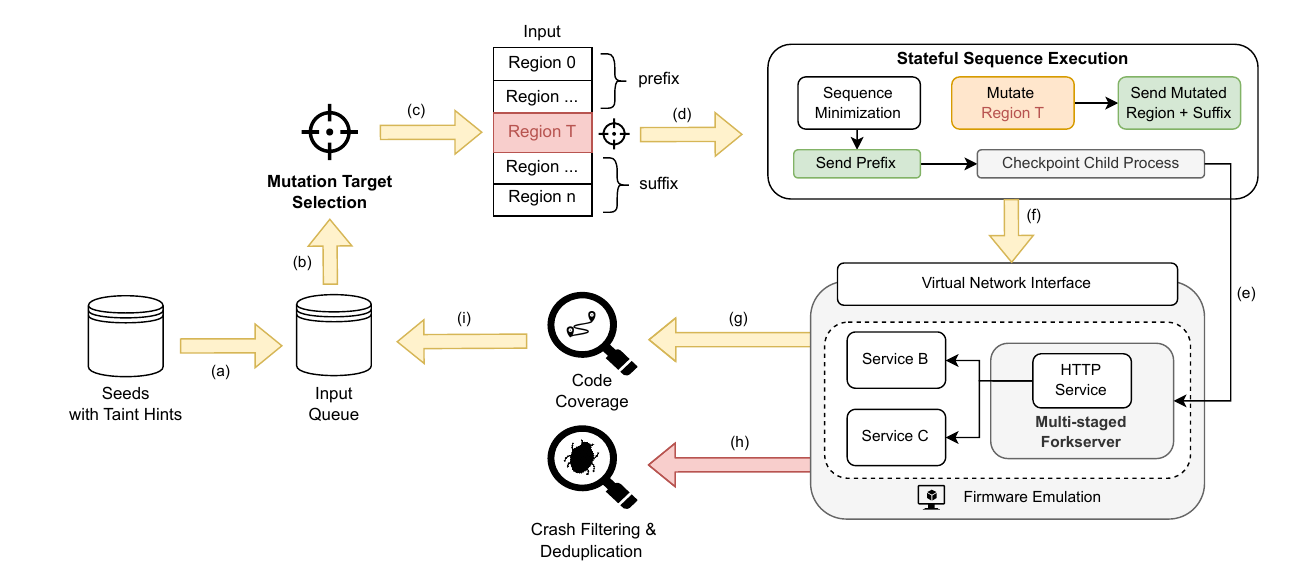}
    \caption{Protocol-Aware Taint-Guided Fuzzing}
    \label{fig:fuzzing_phase}
\end{figure*}

After generating taint hints, \ourtool{} can start the fuzzing phase, whose workflow is depicted in Figure~\ref{fig:fuzzing_phase}. First, the seeds and their related taint hints are imported into the input queue (step a), then an input is picked (step b), determining the fuzzing strategy to consider for the input: {\em state-aware} or {\em taint-guided}. The first one is inspired by protocol-aware fuzzing, while the latter is driven by taint hints (Section~\ref{sssec:taint-hints-heuristic}). Regardless of the strategy, during {\em Mutation Target Selection}, the input is split (step c) into a prefix, a target region T, and a suffix. Then, the input is moved to {\em Stateful Sequence Execution} (step d), where the target region T is mutated according to the strategy and the prefix is replayed (step f), adopting a multi-staged forkserver to amortize the cost of future fuzzing attempts from the same prefix (step e). At this point, the mutated region T and the suffix are replayed (step f). During firmware execution, \ourtool{} saves mutated inputs that lead to crashes for later analysis (step h) and inserts non-crashing inputs into the queue when they achieve improved code coverage (step g). Additional fuzzing attempts with different mutations are then considered over the current input, repeating steps d-i several times. When sufficient effort has been spent mutating an input, the cycle (steps b-i) is repeated by considering other inputs and switching between the two fuzzing strategy. This cycle continues until the time budget for the fuzzing session is fully consumed. The remainder of this section provides detailed descriptions of these steps.

\subsubsection{Input Queue Scheduling}
\label{sssec:queue_scheduling}

The scheduler alternates the two fuzzing strategies to balance \emph{state-aware} exploration with \emph{taint-guided} mutations, ensuring neither mode starves the other and that both receive comparable fuzzing budget during a campaign.

When the state-aware strategy is active, \ourtool{} follows the traditional scheduling in protocol-aware fuzzing (Section~\ref{sec:background}): \ourtool{} maintains a queue of inputs (both seeds and newly generated mutants), grouping them by the server state reached and choosing randomly within each group, following a round-robin policy between groups.

When the taint-guided strategy is active, seeds with taint hints are processed using a round-robin scheme. For each seed, the scheduler retrieves the next unused \emph{taint hint} and schedules a round of fuzzing attempts targeting it. This round-robin approach ensures each seed receives an equal share of fuzzing energy and guarantees that all \emph{taint hints} are eventually explored. When the scheduler reaches the last enqueued seed, it wraps around to the first seed and continues the cycle. For seeds containing multiple hints, a per-seed pointer tracks progress through the hints and a single hint is processed within each queue cycle, maintaining fairness across the entire seed corpus.

\subsubsection{Mutation Target Selection}
\label{sssec:mutation-target}

The {\em Mutation Target Selection} phase decides which part of the input should be mutated. 

With the {\em state-aware} strategy, \ourtool{} adopts state-aware target selection from protocol-aware fuzzing (Section~\ref{ssec:protocol-fuzzing}), which relies on the \textit{Session-Aware HTTP Request Modeling} (see Section~\ref{sssec:request-modeling}) to select the subsequence of requests that corresponds to a particular server state.

With the {\em taint-guided} strategy, the mutation target is chosen according to the taint hint suggested by the input scheduler. Each taint hint specifies a mutation window as a triple ({\tt target}, {\tt offset}, {\tt len}), where {\tt target} is the target region to mutate in the input, {\tt offset} is the starting offset for the mutation, and {\tt len} is the maximum number of bytes to mutate.

\subsubsection{Stateful Sequence Execution}
\label{sssec:sequence-processing}

The \textit{Stateful Sequence Execution} phase performs mutations according to the fuzzing strategy against a chosen input and replays the interaction against the target HTTP service of the firmware, while ensuring each mutated input runs in the correct protocol state and minimizing the replay execution cost. These goals are achieved through a few key ideas that we now review.

\smallskip
{\em Sequence Minimization.} When the target selection is driven by taint hints, \ourtool{} performs a \textit{Sequence Minimization} procedure, aimed at removing any regions before the target region T that do not influence T, keeping only the minimal set of regions in the prefix that are required to preserve data- and filesystem-driven dependencies.

\smallskip
{\em Input mutation.} \ourtool{} adopts two complementary mutation strategies: deterministic bit-flips and havoc stage. Bit-flips are confined to the mutation window selected by the \emph{Mutation Target Selection} and do not alter region length or overall message layout. Both single- and multi-bit flips are used to corrupt token contents, numeric fields, or header bytes while preserving offsets. The havoc stage composes several stacked mutations into each single trial, including value overwrites, small arithmetic operations, block copy/clone, byte-range insertions/deletions, and region overwrites. This allows \ourtool{} to cover both fine-grained corruptions and broader structural distortions of the input.
The same bit-flip and havoc mutations are applied under both taint-guided and state-aware fuzzing strategies. However, one subset of havoc edits, i.e., the region-level operations that change the number of regions in the sequence (region replacement, insertion, duplication, and deletion), are disabled during taint-guided fuzzing. These edits alter the prefix/suffix structure of the request sequence, thereby shifting offsets and breaking the alignment required for taint hints, which assume the original number of regions. By contrast, state-aware mutations are resilient to these shifts, since they do not rely on precise region-preservation semantics.

\smallskip
{\em Multi-Staged Forkserver.} \ourtool{} performs multiple fuzzing attempts on each input. To amortize the replay cost across these attempts, \ourtool{} devises a \emph{Multi-Staged Forkserver}. In particular, like prior works, it exploits a first-stage forkserver reaching the initial state of the HTTP service, which is used when fuzzing an input for the first time. However, during the first fuzzing attempt over an input, the prefix is replayed and a second-stage forkserver is used to cache this intermediate execution state. Any subsequent fuzzing attempt related to the same input can exploit the second-stage forkserver to reduce replay execution cost.

\subsubsection{Whole-system Code Coverage Tracing}
\label{sssec:code_coverage_trace}

Code coverage is a major feedback mechanism in fuzzing (see Section~\ref{ssec:software_fuzzing}). A single code coverage bitmap for all processes, ignoring that the same (user-space) code address can be used by different processes, conflates identical virtual PCs that actually belong to different processes, producing many spurious collisions. To mitigate such a problem, \ourtool{} tracks code module loading, recording for each process and each module the inode related to the loaded binary file, the start address, and the length of the text section. When a basic block executes, the code coverage bitmap is updated according to an index that depends on a combination of the normalized\footnote{To cope with address layout randomization.} program counter and the inode related to the module of currently running process. Moreover, code coverage of kernel space and common shared libraries is filtered out to avoid noisy and non-informative coverage.


\subsubsection{Crash Filtering \& Deduplication}
\label{sssec:crash-detection}
Detecting and correctly triaging crashes in full-system fuzzing requires distinguishing true input-induced faults from unrelated emulation errors. To this end, a dedicated \textit{Crash Filtering \& Deduplication} pipeline was implemented.

\ourtool{} identifies any user-space fatal fault inside the firmware emulation environment. Indeed, the Linux kernel used for emulation is instrumented to invoke a stub in \texttt{do\_coredump()} whenever a process receives a fatal signal (e.g., {\tt SIGSEGV}, {\tt SIGBUS}). This guarantees that both long-running daemons and short CGI helpers are trapped immediately at the point of failure, without relying on delayed user-space detection.
Moreover, throughout execution, \ourtool{} maintains a circular buffer of the last \(N\) executed basic blocks for each distinct process. Upon trap, the content of the buffer is used as the {\em crash fingerprint}. An example of crash fingerprints is shown below:
\begin{verbatim}
=== Fingerprint #0 ===
Faulty process: setup.cgi
  [00] inode: 24746, pc: 0x00012244, module: setup.cgi
  [01] inode: 24746, pc: 0x00008384, module: setup.cgi
  [02] inode: 24746, pc: 0x000106d8, module: setup.cgi
  [03] inode: 24746, pc: 0x000106cc, module: setup.cgi
  [04] inode: 24746, pc: 0x00011d10, module: setup.cgi

=== Fingerprint #1 ===
Faulty process: potcounter
  [00] inode: 24728, pc: 0x0000070c, module: potcounter
  [01] inode: 24728, pc: 0x000006d0, module: potcounter
  [02] inode: 24728, pc: 0x000007c4, module: potcounter
  [03] inode: 24728, pc: 0x0000099c, module: potcounter
  [04] inode: 24728, pc: 0x00000a04, module: potcounter
\end{verbatim}

As illustrated above, different processes may fault during a fuzzing attempt. In \texttt{Fingerprint~\#0}, the failure arises in the \texttt{setup.cgi} process, while in \texttt{Fingerprint~\#1} it occurs in the \texttt{potcounter} process. This highlights a key aspect of full-system fuzzing: a single mutated input may concurrently trigger distinct failures across multiple services running inside the emulated firmware. By fingerprinting each process’s recent basic block history independently, the fuzzer is able to detect, classify, and deduplicate such multi-process crashes.
Two crashes are thus considered equivalent when their \emph{crash fingerprints} match, even across different fuzzing iterations. Unlike stack traces, which are often unavailable or unreliable in stripped firmware binaries, the histories of recently executed basic block are lightweight and consistently available.

One major complication is that some processes may fault deterministically during system initialization (e.g., due to missing peripherals or incomplete one-time setup). Such crashes are unrelated to input mutations and must not be reported as bugs. To filter them, a \textit{dry-run phase} is performed before fuzzing: every seed in the initial corpus (assumed by definition unfaulty) is executed once, and all crash fingerprints observed during this phase are collected into an ignore list that is used to suppress spurious crashes. For example, the \texttt{potcounter} fault shown in \texttt{Fingerprint~\#1} above occurs consistently even during dry runs, and is therefore classified as an emulation artifact rather than a true crash.

However, this fingerprinting strategy may not always be sufficient for perfect deduplication. Indeed, two distinct bugs that follow different execution traces may converge on the same recent basic block history would lead to identical fingerprints. In such cases, different crashes are not reported separately but merged. This limitation is accepted as a trade-off: by restricting attention to the immediate control-flow execution history preceding the fault, the approach often provides good strategy to cheaply cluster duplicated instances of the same bug.

\subsection{Discussion}

Table~\ref{table:feature_comparison} compares the features of \ourtool{} with TriforceAFL and AFLNet, two prominent fuzzing frameworks relevant to our context.

Request modeling is crucial for achieving reproducible interactions in stateful applications. TriforceAFL is unaware of the underlying protocol logic, blindly breaking interactions into different inputs. AFLNet implements a session-aware mode but still depends on unreliable timeout heuristics to delimit responses. In contrast, \ourtool{} exploits Session-Aware HTTP Request Modeling (Section~\ref{sssec:request-modeling}) to reconstruct complete request/response exchanges using protocol-semantic rules rather than timing.

Sequence Minimization is a capability unique to \ourtool{} (Section~\ref{sssec:queue_scheduling}). It filters out irrelevant requests from the original interactions without harming intra- and inter-service dependencies. This minimization reduces replay cost, potentially increasing the number of fuzzing attempts.

Mutation strategies represent another essential difference. TriforceAFL integrates classical mutations from AFL, which do not always work well with stateful applications. AFLNet devises protocol-aware mutations that can drastically improve fuzzing effectiveness; however, it is not aware of which parts of the inputs affect the firmware execution. \ourtool{} complements protocol-aware mutations with taint-guided mutations (Section~\ref{sssec:sequence-processing}), prioritizing inputs that impacted a large number of basic blocks and achieving finer-grained guidance than stateful heuristics alone.

Code coverage tracking also shows a fundamental divergence. TriforceAFL and AFLNet do not discriminate between different processes, experiencing an excessive number of collisions in their coverage bitmap. \ourtool{} is process-aware and attempts to minimize bitmap collisions (Section~\ref{sssec:code_coverage_trace}).

Fuzzing efficiency is improved by \ourtool{}'s Multi-Staged Forkserver (Section~\ref{sssec:mutation-target}), which supports forking even to intermediate execution states. This amortizes the cost of long input sequence replay and enables many mutation trials from the same intermediate execution state. AFLNet and TriforceAFL only support traditional single-stage forkservers.

Crash triage presents another key distinction. TriforceAFL and AFLNet treat faulty inputs as distinct crashes when their coverage bitmaps do not perfectly match. While this strategy is simple, it often fails to deduplicate crashes: small, early, and irrelevant divergences during execution may lead to retaining different crashing inputs even when they relate to the same bug. While state-of-the-art user-mode fuzzers often exploit stack traces to improve deduplication, system-mode emulation makes stack trace construction non-trivial and costly. \ourtool{} performs crash triaging in distinct phases (Section~\ref{sssec:crash-detection}): filtering and deduplication. In the first one, it discards spurious failures that do not depend on the mutated input. In the second one, it deduplicates crashes based on recent basic block execution history, which can significantly help group crashes related to the same bug.



\begin{table*}[ht]
\renewcommand{\arraystretch}{1.1}
\setlength{\tabcolsep}{4pt}
\centering
\begin{tabular}{|l||c|c|c|}

\hline
{\sc Feature} & {\sc TriforceAFL}~\cite{triforceafl} & {\sc AFLNet}~\cite{pham2020aflnet} & \ourtool{} \\
\toprule
\hline
Request Modeling & No & Timeout & Session Aware \\
\hline
Mutations & AFL & State Aware & State Aware + Taint Guided \\
\hline
Sequence Minimization & No & No & Yes \\
\hline
Forkserver & Single Stage & Single Stage & Multi Stage \\
\hline
Code Coverage Tracing & Single process & Single process & Multiple processes \\
\hline
Crash Filtering & No & No & Spurious Crashes \\
\hline
Crash Deduplication & Coverage Bitmap & Coverage Bitmap & BB History \\
\hline
\end{tabular}
\caption{Comparison of core features across {\sc TriforceAFL}, {\sc AFLNet}, and \ourtool{}.}
\label{table:feature_comparison}
\end{table*}


\section{Implementation}
\label{sec:implementation}

We developed \ourtool{} using approximately 1,100 lines of Python and 4,000 lines of C/C++ code\footnote{\url{https://github.com/alessioizzillo/STAFF}}. This section provides several details behind the implementation of \ourtool{}, highlighting the key changes we made on top of existing rehosting, taint, and fuzzing frameworks, and the optimizations  necessary to keep the system scalable.



\subsection{Whole-System Taint Analysis}
\label{ssec:decaf_optimizations}

A major effort during the development of \ourtool{} was targeted at improving the whole-system taint analysis layer based on QEMU~\cite{qemu} from DECAF++~\cite{davanian2019decaf++}. To catch taint as soon as values flow into userland, \ourtool{} injects taint at the kernel-level write completion handler by instrumenting \texttt{address\_space\_write\_con}- \texttt{tinue()} from QEMU. This guarantees that values mapped into user memory from MMIO, kernel mediation, or other kernel sources are tainted immediately upon kernel-to-user transfer. Since QEMU emits multi-byte load and store operations, \ourtool{} fragments them into byte-level sub-operations, emitting one tainted memory-access event per byte. We further added explicit support for unaligned partial-word semantics (SWL/SWR) and injected taint operations directly into the translation-block generation path through \texttt{tcg\_emit\_op()}. These steps reduced missed flows and significantly increased the fidelity of our byte annotations.

To capture filesystem operations (Section~\ref{ssec:streaming}), \ourtool{} instruments primary read and write syscalls, recording writer-reader relationships between regions. The resulting dependency map is filtered to ignore irrelevant files such as shared libraries or timezone files. This filtering mitigates over-dependencies and produces minimized sequences that remain compact and focused on causally relevant writers. In practice, this choice substantially reduced noisy prefix and suffix retention without missing genuine dependencies.

\subsection{Byte Annotation Heuristic Optimizations}
\label{ssec:byte_ann_opt}

Two components of our implementation are particularly sensitive to combinatorial blowup: the trie construction performed by the procedure {\em IndexRegionByteSequencesInTrie} in Algorithm~\ref{alg:byte-annotation-emilio} and the trie-based subsequence matching used to infer inter-region dependencies in Algorithm~\ref{alg:annotate-subregion-emilio}. We applied a number of optimizations to keep memory and CPU costs bounded while maximizing precision.

During trie construction, procedure {\em IndexRegionByteSequencesInTrie} considers region subsequences up to a capped maximum length. Moreover, when the memory usage during trie construction exceeds a user-defined threshold, the trie construction is aborted and re-attempted with a halved maximum length.

During trie-based subsequence matching, Algorithm~\ref{alg:byte-annotation-emilio} is quadratic with respect to $|E|$ and may quickly become intractable. We bound $\ell$ to an interval such that extremely short subsequences, which tend to match many positions, are avoided, while overly long ones are capped. A further optimization is to stop as soon as the first unique trie match is found (line~\ref{line:check-match}). This early exit avoids unnecessary exploration of additional subsequences once a reliable mapping has already been discovered.


\subsection{Taint Hints Ordering}
\label{subsec:taint-hint-optimization}

The \emph{SortFilterHints} procedure (Algorithm~\ref{alg:filter-hints}) performs the length-based ordering and the overlap filtering as described in Section~\ref{sssec:taint-hints-heuristic}. In our implementation, we further sort the filtered hints at the end of the procedure to later guide the fuzzing scheduler: \ourtool{} prioritizes hints covering more basic blocks, with the intuition that mutating their bytes may impact a large number of basic blocks and possibly expose bugs faster. By applying this ordering we noticed that it may help reduce the time-to-exposure (TTE) of crashes on some firmware images.

\section{Evaluation}
\label{sec:evaluation}

This section evaluates the effectiveness of \ourtool{} when considering real-world firmware images. 
Our goal is to validate the ability of \ourtool{} in discovering real-world firmware bugs and the efficiency of the proposed ideas, comparing our implementation with the state-of-the-art.
In more detail, we summarize the evaluation of \ourtool{} around the following research questions:
\begin{itemize}

\item \textbf{RQ1:} How precise is the \emph{Byte Annotation Heuristic} compared to traditional taint analysis, and what is the impact of \emph{Sequence Minimization}?

\item \textbf{RQ2:} Does \ourtool{} improve bug discovery in terms of number of bugs found, crash reproducibility, and time-to-exposure, compared to the state of the art?

\item \textbf{RQ3:} Do the bugs found by \ourtool{} involve inter-binary or multi-request interactions?

\item \textbf{RQ4:} Why is \ourtool{} able to discover these bugs?

\end{itemize}
Before presenting our experimental results, we describe our experimental setup, including the considered dataset.


\subsection{Experimental Setup}
\label{ssec:dataset}

\smallskip
\noindent\textit{Experimental Environment.}
Experiments ran on dual-socket Intel Xeon Gold 6238R CPU (56 cores). To ensure reproducibility and safe parallelism, the FirmAE image-preparation pipeline was extended to emit per-tool artifacts and package each firmware run into isolated Docker environments: per-tool databases replaced the previous monolithic PostgreSQL store to remove global contention, and each fuzzer instance executes in its own container with an independent network namespace and resource limits.

Selected runtime parameters were chosen to balance fidelity, reproducibility, and throughput. The minimum subsequence length $\ell$ for the \emph{Intra- and Inter-Service Dependency Analysis} was set to 3 to filter out 1- and 2-byte trivially short matches that produce most ambiguous trie hits. Tracing of shared libraries was disabled to avoid noisy coverage from commonly-invoked library code and focus instrumentation on application binaries. Each fuzzing trial ran for 24~h (per-trial budget), and individual test cases were bounded by a 150~s per-input timeout to prevent hung or excessively slow emulations from stalling the campaign. The running time of \emph{Intra- and Inter-Service Dependency Analysis} was excluded from \ourtool{}'s 24~h fuzzing budget to ensure fair comparison. Coverage bitmap sizes were tuned per firmware in the range $2^{22}$--$2^{25}$ to reduce collisions for larger code execution while keeping RAM usage reasonable. AFL's timing calibration was disabled and queue shuffling turned off to improve repeatability in our multi-process emulation environment, where scheduling nondeterminism otherwise produces unstable timing baselines. AFLNet inputs are delimited by the four-byte marker \texttt{0x1A1A1A1A} to separate region boundaries reliably. Each mutation was applied up to three times to elicit multiple behaviors from the same edit, and the fuzzing dictionary contained a single long token (1,260 bytes) intended to exercise common lexical fields without introducing many short, extraneous tokens.

\smallskip
\noindent\textit{Dataset.} 
The evaluation corpus (Table~\ref{table:dataset}) is a carefully curated subset of firmware images drawn from the FirmAE project~\cite{kim2020firmae}. It contains routers, range extenders, and IP cameras from multiple vendors, produced by screening a larger collection of images and retaining only those suitable for reproducible, interactive full-system fuzzing. We kept images whose embedded web server was reachable and exposed a usable management interface, where the firmware would boot and initialize without critical failures under emulation, and where user interactions through the web UI were responsive enough to be captured and replayed. Each selected image admits at least one valid web session that can be recorded as a PCAP trace and reliably replayed by our orchestration. We excluded images that require browser-specific behavior, reject automated clients, or use obfuscated/encrypted authentication flows that prevent deterministic replay and taint tracing. We also excluded images that only present static landing pages (e.g., vendor information pages) and do not expose the device management functions that our network-driven fuzzing targets. These selection criteria yield a reproducible, realistic dataset appropriate for comparing stateful, multi-request fuzzing techniques.

\smallskip
{\em Competitors.}
The evaluation compares \ourtool{} against two state-of-the-art baselines: AFLNet~\cite{pham2020aflnet} and TriforceAFL~\cite{triforceafl}. Both baselines were adapted to run in a full-system, multi-process emulation context so that comparisons are fair and apples-to-apples.

AFLNet was ported to system-mode (referred to in the paper as AFLNet$^{\dagger}$) by replacing its original qemu-user workflow with a FirmAE-based system VM that hosts a forkserver. The port includes the harness which coordinates QEMU and the forkservers via signals to spawn/exit worker children that execute mutated inputs. This allowed AFLNet$^{\dagger}$ to exercise full-system behaviour (multiple daemons, inter-process interactions) rather than a single userland process.

TriforceAFL is derived from a full-system FirmAFL variant and was adapted to the multi-process setting by improving its {\tt accept} syscall detection heuristic and multi-process-handling logic so that network-driven requests are recognized reliably in the emulated environment. These changes were necessary to make TriforceAFL operate robustly when multiple processes run concurrently.

To ensure a fair comparison on coverage-driven feedback, both AFLNet$^{\dagger}$ and TriforceAFL were aligned with \ourtool{}’s \emph{Whole-system Code Coverage Tracing} (see Section~\ref{sssec:code_coverage_trace}). This unification prevents cross-process pollution and reduces bitmap collisions, so differences in results reflect methodological choices (taint guidance, sequence minimization, checkpointing) rather than artifacts of mismatched tracing implementations.

\begin{table*}[ht!]
\renewcommand{\arraystretch}{1.1}
\setlength{\tabcolsep}{4pt}
\centering
\begin{tabular}{|l||c|c|c|c|c|}
\hline
{\sc Firmware} & {\sc Vendor} & {\sc Version} & {\sc Type} & {\sc Architecture} & {\sc Captured Sessions} \\
\toprule
\hline
RT\_N10U & \multirow{2}{*}{ASUS} & 30043763754 & Router & MIPSEL & 4 \\
\cline{1-1}\cline{3-6}
RT\_N53 &  & 30043763754 & Router & MIPSEL & 4 \\
\hline
DAP-2310 & \multirow{3}{*}{D-Link} & v1.00\_o772 & Router & MIPSEB & 4 \\
\cline{1-1}\cline{3-6}
DIR-300 &  & v1.03\_7c & Router & MIPSEB & 4 \\
\cline{1-1}\cline{3-6}
DIR-815A1 &  & FW104b03 & Router & MIPSEL & 5 \\
\hline
RE1000 & \multirow{2}{*}{Linksys} & 1.0.02.001\_US\_20120214 & Range Extender & MIPSEL & 2 \\
\cline{1-1}\cline{3-6}
WRT320N &  & 1.0.05.002\_20110331 & Router & MIPSEL & 4 \\
\hline
DGN3500 & \multirow{3}{*}{Netgear} & V1.1.00.30\_NA & Router & MIPSEB & 5 \\
\cline{1-1}\cline{3-6}
DGND3300 & & V1.1.00.22\_NA & Router & MIPSEB & 5 \\
\cline{1-1}\cline{3-6}
JNR3210 & & V1.1.0.14 & Router & MIPSEB & 4 \\
\hline
Archer C2 & \multirow{2}{*}{TP-Link} & v1\_160128 & Router & MIPSEL & 4 \\
\cline{1-1}\cline{3-6}
TL-WPA8630 & & V2\_171011 & Range Extender & MIPSEB & 4 \\
\hline
TV-IP121WN & \multirow{3}{*}{TRENDnet} & 1.2.2 & IP Camera & MIPSEB & 4 \\
\cline{1-1}\cline{3-6}
TV-IP651WI & & V1\_1.07.01 & IP Camera & MIPSEL & 4 \\
\cline{1-1}\cline{3-6}
TEW-652BRU &  & 1.00b12 & Router & MIPSEB & 4 \\
\hline
\end{tabular}
\caption{Dataset of firmware images used in the evaluation.}
\label{table:dataset}
\end{table*}

\subsection{RQ1: Byte Annotation Precision}
\label{ssec:rq1}

\begin{table}[ht!]
\renewcommand{\arraystretch}{1.1}
\setlength{\tabcolsep}{3pt}
\centering
\begin{tabular}{|l||c|c|c|}
\hline
\multirow{2}{*}{{\sc Firmware}} & \multicolumn{1}{c|}{{\sc Byte Annotation}} & \multicolumn{2}{c|}{{\sc Seq. Minimization}} \\
\cline{2-4}
& {\sc Precision} & {\sc Speedup} & {\sc Precision} \\
\toprule
\hline
RT-N10U & 0.71 & 11.56$\times$ & 0.92 \\
\hline
RT-N53 & 0.87 & 8.28$\times$ & 0.87 \\
\hline
DIR-815 & 0.94 & 1.45$\times$ & 0.97 \\
\hline
DAP-2310 & 0.92 & 1.62$\times$ & 0.93 \\
\hline
DIR-300 & 0.88 & 4.61$\times$ & 0.91 \\
\hline
FW\_RE1000 & 0.99 & 4.96$\times$ & 0.91 \\
\hline
WRT320N & 0.86 & 5.41$\times$ & 0.9 \\
\hline
DGN3500 & 0.93 & 9.29$\times$ & 0.89 \\
\hline
DGND3300 & 0.9 & 4.29$\times$ & 0.73 \\
\hline
JNR3210 & 0.88 & 10.22$\times$ & 0.95 \\
\hline
Archer C2 & 0.73 & 10.78$\times$ & 0.84 \\
\hline
TL-WPA8630 & 0.91 & 16.68$\times$ & 0.92 \\
\hline
TV-IP121WN & 0.91 & 14.29$\times$ & 0.94 \\
\hline
TV-IP651WI & 0.46 & 1.47$\times$ & 0.95 \\
\hline
TEW-652BRU & 0.91 & 6.95$\times$ & 0.9 \\
\hline\hline
{\sc Geo Mean} & 0.84 & 5.82$\times$ & 0.9 \\ 
\hline
\end{tabular}
\caption{Precision of the byte annotations before and after \emph{Sequence Minimization}, and speedup of replaying the minimized sequence compared to the original one.}
\label{tab:staff_perf}
\end{table}

An essential trait of \ourtool{} is byte annotations, which serve as the backbone for taint hints, which in turn allow \ourtool{} to perform targeted mutations based on firmware execution behavior. These byte annotations are obtained by combining a coarse-grained dynamic taint analysis (using a single taint source) with a value matching strategy designed to refine taints and recover fine-grained byte-level dependencies. This approach has been designed as a nice trade-off between accuracy and scalability compared to traditional dynamic taint analysis, which may assign distinct taint sources to each input byte but would suffer from scalability limitations. However, this raises a natural question: {\em how precise is the approximated taint analysis from} \ourtool{} {\em compared to full taint analysis with byte-level taint sources?} In the remainder of this section, we aim to provide an answer, considering also the sequence minimization phase that further affects the taints.

In more detail, we designed an experiment that, given a taint hint {\em h} generated by \ourtool{} for the input bytes $[h.\texttt{offset}, h.\texttt{offset}+h.\texttt{len}-1]$, considers the related byte annotations that lead to hint creation, collecting all basic blocks $\textit{PCs}_{\text{\ourtool{}}}$ impacted by the considered input bytes according to the byte annotations. We then run a dynamic taint analysis, marking each input byte in $[h.\texttt{offset}, h.\texttt{offset}+h.\texttt{len}-1]$ with distinct taint sources, collecting all basic blocks $\textit{PCs}_{\text{DTA}}$ impacted by the considered input bytes according to the fine-grained dynamic taint analysis. To account for execution variability, we repeat the experiments five times and merge basic block lists from different runs. To avoid bias in the taint hint choice, we consider 10 randomly picked taint hints generated during a run over a randomly chosen seed input. Finally, we compute the mean precision across the 10 taints, defined as:
$$
  \text{precision} = \frac{|\textit{PCs}_{\text{\ourtool{}}} \cap \textit{PCs}_{\text{DTA}}|}{|\textit{PCs}_{\text{\ourtool{}}}|}
$$
Since sequence minimization may filter out byte annotations, we compute the precision before and after minimization.



Table~\ref{tab:staff_perf} summarizes the results of the experiment. The mean precision of the byte annotations, before applying sequence minimization, is high for most firmware images, suggesting that byte annotations correctly represent, in the majority of cases, real taint dependencies between input bytes and influenced basic blocks and therefore are unlikely to be spurious associations.

When considering the impact of sequence minimization, the precision of the surviving byte annotations remains significantly high, with an increase in precision in several firmware images compared to the byte annotations before minimization. This result suggests that the taint hints considered by \ourtool{} faithfully represent taint dependencies. Moreover, the filtering is clearly justified by significant execution speedups during the replay of the minimized sequences compared to the original ones, supporting our design choice.

Finally, we comment on the fact that the experiments do not focus on reporting the recall, i.e., how many taint dependencies are ignored by \ourtool{}. This choice stems from the fact that computing the full set of dependencies is impractical and that, by design, \ourtool{} discards some taint hints as a design choice to keep the approach scalable. To mitigate the lack of this evaluation, in the next research questions, we consider the practical value of the taint hints considered by \ourtool{} when analyzing crashes and bugs, which, in the end, is the ultimate goal of \ourtool{}.

\begin{answer}{1}
The \emph{Byte Annotation Heuristic} provides a precise approximation of full per-byte dynamic taint analysis. \emph{Sequence Minimization} preserves high precision (mean 0.9 across firmware images) while delivering substantial execution speedup (mean $5.82\times$ improvement) during input replay, significantly enhancing the approach's scalability.
\end{answer}

\subsection{RQ2: Fuzzing Effectiveness}
\label{ssec:rq2}

\begin{table}[ht]
\centering
\renewcommand{\arraystretch}{1.1}
\setlength{\tabcolsep}{4pt}
\begin{tabular}{|l||c|c|c|}
\hline
{\sc Firmware} & {\sc TriforceAFL} & {\sc AFLNet}$^{\dagger}$ & \ourtool{} \\
\toprule
\hline
DGN3500 & 1.0 & 4.0 & 5.2 \\
\hline
DGND3300 & 1.4 & 3.8 & 5.2 \\
\hline
TV-IP121WN & 0.0 & 1.0 & 1.0 \\
\hline
TV-IP651WI & 0.0 & 0.0 & 1.0 \\
\hline
WRT320N & 2.8 & 1.4 & 4.8 \\
\hline
JNR3210 & 2.0 & 2.0 & 3.6 \\
\hline
TL-WPA8630 & 1.0 & 2.0 & 2.0 \\
\hline
DAP-2310 & 0.0 & 1.0 & 3.2 \\
\hline
DIR-300 & 0.0 & 1.0 & 4.4 \\
\hline
\end{tabular}
\caption{Average number of unique bugs found during 24-hour runs.}
\label{table:number_bugs}
\end{table}

\begin{table*}[ht]
\centering
\renewcommand{\arraystretch}{1.12}
\setlength{\tabcolsep}{3pt}
\begin{tabular}{|l||c|c|c|c|c|c|c|c|c|c|}
\hline
\multirow{2}{*}{{\sc Firmware}} & \multirow{2}{*}{{\sc Binary}} & \multirow{2}{*}{{\sc Function}} & {\sc Bug} & \multicolumn{2}{c|}{{\sc TriforceAFL}} & \multicolumn{2}{c|}{{\sc AFLNet$^{\dagger}$}} & \multicolumn{2}{c|}{{\sc \ourtool{}}} \\
\cline{5-10}
 &  &  &  {\sc Category} & {\sc \#} & {\sc TTE} & {\sc \#} & {\sc TTE} & {\sc \#} & {\sc TTE} \\
\hline\hline
\multirow{10}{*}{DGN3500} & \multirow{1}{*}{\texttt{mini\_httpd}} & \texttt{FUN\_00410e6c} & OIB & 5 & 4h25m & 5 & 0h58m & 5 & 2h22m \\
\cline{3-10}
\cline{2-10}
 & \multirow{1}{*}{\texttt{rc}} & \texttt{get\_sd\_info} & MII & 0 &  & 4 & 3h32m & 5 & 5h34m \\
\cline{3-10}
\cline{2-10}
 & \multirow{8}{*}{\texttt{setup.cgi}} & \texttt{addkeyword} & MID & 0 &  & 0 &  & 3 & 16h39m \\
\cline{3-10}
 &  & \texttt{delete} & MID & 0 &  & 2 & 7h1m & 0 &  \\
\cline{3-10}
 &  & \texttt{find\_val} & MID & 0 &  & 2 & 10h18m & 1 & 22h31m \\
\cline{3-10}
 &  & \texttt{html\_parser} & MID & 0 &  & 5 & 1h33m & 5 & 1h3m \\
\cline{3-10}
 &  & \texttt{save} & MID & 0 &  & 0 &  & 2 & 19h27m \\
\cline{3-10}
 &  & \texttt{set\_SRouteMetric} & MID & 0 &  & 0 &  & 5 & 6h3m \\
\cline{3-10}
 &  & \texttt{set\_TimeZone} & MID & 0 &  & 1 & 4h17m & 0 &  \\
\cline{3-10}
 &  & \texttt{set\_rule\_out} & MID & 0 &  & 1 & 10h50m & 0 &  \\
\cline{3-10}
\cline{2-10}
\cline{1-1}
\multirow{13}{*}{DGND3300} & \multirow{2}{*}{\texttt{mini\_httpd}} & \texttt{FUN\_004036e8} & OIB & 2 & 0h34m & 0 &  & 0 &  \\
\cline{3-10}
 &  & \texttt{FUN\_0040427c} & OIB & 5 & 0h56m & 2 & 11h4m & 2 & 19h29m \\
\cline{3-10}
\cline{2-10}
 & \multirow{11}{*}{\texttt{setup.cgi}} & \texttt{addkeyword} & MID & 0 &  & 1 & 13h19m & 3 & 12h53m \\
\cline{3-10}
 &  & \texttt{del\_list} & MID & 0 &  & 3 & 2h20m & 3 & 13h56m \\
\cline{3-10}
 &  & \texttt{find\_val} & MID & 0 &  & 0 &  & 1 & 22h27m \\
\cline{3-10}
 &  & \texttt{get\_WAN\_ipType} & MID & 0 &  & 1 & 17h51m & 1 & 18h47m \\
\cline{3-10}
 &  & \texttt{html\_parser} & MID & 0 &  & 5 & 0h35m & 5 & 0h57m \\
\cline{3-10}
 &  & \texttt{save} & MID & 0 &  & 2 & 14h29m & 2 & 5h19m \\
\cline{3-10}
 &  & \texttt{set\_SRouteMetric} & MID & 0 &  & 0 &  & 5 & 4h15m \\
\cline{3-10}
 &  & \texttt{set\_TimeZone} & MID & 0 &  & 1 & 0h19m & 1 & 17h41m \\
\cline{3-10}
 &  & \texttt{set\_WAN\_ipType} & MID & 0 &  & 1 & 4h37m & 0 &  \\
\cline{3-10}
 &  & \texttt{set\_rule\_in} & MID & 0 &  & 2 & 7h44m & 0 &  \\
\cline{3-10}
 &  & \texttt{set\_rule\_out} & MID & 0 &  & 1 & 0h45m & 3 & 9h47m \\
\cline{3-10}
\cline{2-10}
\cline{1-1}
\multirow{2}{*}{TV-IP121WN} & \multirow{1}{*}{\texttt{network.cgi}} & \texttt{main} & OID & 0 &  & 0 &  & 5 & 0h33m \\
\cline{3-10}
\cline{2-10}
 & \multirow{1}{*}{\texttt{view.cgi}} & \texttt{main} & OID & 0 &  & 1 & 14h59m & 0 &  \\
\cline{3-10}
\cline{2-10}
\cline{1-1}
\multirow{1}{*}{TV-IP651WI} & \multirow{1}{*}{\texttt{alphapd}} & \texttt{websCgibinProcessor} & MID & 0 &  & 0 &  & 3 & 1h11m \\
\cline{3-10}
\cline{2-10}
\cline{1-1}
\multirow{5}{*}{WRT320N} & \multirow{5}{*}{\texttt{httpd}} & \texttt{get\_cgi} & OIB & 5 & 16h30m & 0 &  & 4 & 4h29m \\
\cline{3-10}
 &  & \texttt{valid\_hwaddr} & OIB & 0 &  & 5 & 4h49m & 5 & 1h14m \\
\cline{3-10}
 &  & \texttt{validate\_cgi} & OIB & 5 & 16h21m & 0 &  & 5 & 2h56m \\
\cline{3-10}
 &  & \texttt{validate\_forward\_single} & OIB & 0 &  & 2 & 6h55m & 5 & 2h17m \\
\cline{3-10}
 &  & \texttt{validate\_merge\_ipaddrs} & OIB & 4 & 16h38m & 0 &  & 5 & 0h52m \\
\cline{3-10}
\cline{2-10}
\cline{1-1}
\multirow{4}{*}{JNR3210} & \multirow{2}{*}{\texttt{mini\_httpd}} & \texttt{FUN\_00403acc} & OIB & 0 &  & 0 &  & 3 & 9h35m \\
\cline{3-10}
 &  & \texttt{FUN\_00405c88} & OIB & 5 & 1h1m & 5 & 0h30m & 5 & 2h25m \\
\cline{3-10}
\cline{2-10}
 & \multirow{2}{*}{\texttt{setup.cgi}} & \texttt{html\_parser} & OID & 5 & 0h34m & 5 & 1h31m & 5 & 7h37m \\
\cline{3-10}
 &  & \texttt{upgrade\_main} & OID & 0 &  & 0 &  & 5 & 4h28m \\
\cline{3-10}
\cline{2-10}
\cline{1-1}
\multirow{5}{*}{TL-WPA8630} & \multirow{4}{*}{\texttt{httpd}} & \texttt{FUN\_004057d0} & OIB & 0 &  & 1 & 2h4m & 5 & 13h52m \\
\cline{3-10}
 &  & \texttt{FUN\_0040edfc} & MIB & 0 &  & 2 & 4h41m & 4 & 10h57m \\
\cline{3-10}
 &  & \texttt{FUN\_0041685c} & MIB & 0 &  & 1 & 4h39m & 0 &  \\
\cline{3-10}
 &  & \texttt{FUN\_00425f90} & OIB & 5 & 0h1m & 0 &  & 0 &  \\
\cline{3-10}
\cline{2-10}
 & \multirow{1}{*}{\texttt{ledschd}} & \texttt{FUN\_00402320} & MII & 0 &  & 2 & 4h13m & 1 & 6h4m \\
\cline{3-10}
\cline{2-10}
\cline{1-1}
\multirow{6}{*}{DAP-2310} & \multirow{3}{*}{\texttt{atp}} & \texttt{ExeShell} & MID & 0 &  & 0 &  & 2 & 8h56m \\
\cline{3-10}
 &  & \texttt{redirect\_htm} & MID & 0 &  & 0 &  & 5 & 3h4m \\
\cline{3-10}
 &  & \texttt{rpsubname} & MID & 0 &  & 0 &  & 1 & 23h20m \\
\cline{3-10}
\cline{2-10}
 & \multirow{1}{*}{\texttt{udhcpd}} & \texttt{FUN\_00403448} & MII & 0 &  & 2 & 6h58m & 2 & 3h23m \\
\cline{3-10}
\cline{2-10}
 & \multirow{2}{*}{\texttt{xmldb}} & \texttt{FUN\_00408618} & MII & 0 &  & 0 &  & 5 & 2h31m \\
\cline{3-10}
 &  & \texttt{FUN\_00408b50} & MII & 0 &  & 0 &  & 1 & 17h8m \\
\cline{3-10}
\cline{2-10}
\cline{1-1}
\multirow{5}{*}{DIR-300} & \multirow{3}{*}{\texttt{atp}} & \texttt{ExeShell} & MID & 0 &  & 0 &  & 3 & 9h5m \\
\cline{3-10}
 &  & \texttt{do\_xgi} & MID & 0 &  & 0 &  & 5 & 1h43m \\
\cline{3-10}
 &  & \texttt{redirect\_htm} & MID & 0 &  & 1 & 11h34m & 5 & 1h19m \\
\cline{3-10}
\cline{2-10}
 & \multirow{1}{*}{\texttt{udhcpd}} & \texttt{FUN\_00402f70} & MII & 0 &  & 0 &  & 4 & 4h17m \\
\cline{3-10}
\cline{2-10}
 & \multirow{1}{*}{\texttt{xmldb}} & \texttt{FUN\_0040b7e4} & MII & 0 &  & 0 &  & 5 & 1h20m \\
\cline{3-10}
\cline{2-10}
\cline{1-1}
\hline
\end{tabular}
\caption{Unique bugs found by the three approaches.}
\label{table:tte_bugs}
\end{table*}

The ultimate goal of \ourtool{} is finding bugs in firmware images. Hence, in this section, we investigate whether the ideas proposed by \ourtool{} can actually make it identify more crashes than TriforceAFL and AFLNet$^{\dagger}$, two existing state-of-the-art system-mode firmware fuzzing solutions. To this aim, we executed the different fuzzing frameworks over our dataset for 24 hours, repeating each run five times.

Table~\ref{table:number_bugs} presents the average number of unique bugs discovered by the different approaches during the five runs for each firmware from our dataset\footnote{Firmware images without a crash are omitted.}. \ourtool{} can consistently identify more bugs than the competitors, finding on average across the 15 firmware $2.03$ bugs, compared to $1.08$ and $0.55$ for AFLNet$^{\dagger}$ and TriforceAFL, respectively. These results suggest that the features of \ourtool{} can indeed make a difference in bug discovery.

To have a more fine-grained investigation of the unique bugs discovered by the different tools, Table~\ref{table:tte_bugs} refines the analysis of the experimental results by reporting for each firmware, for each unique bug, and for each approach, the Time-To-Exposure (TTE) of the bug and the number of times the bug was identified during the five runs. Furthermore, in the table, each bug is identified by the binary and function where it is located and a categorization. In particular, we consider the following bug categories:

\begin{itemize}

\item[{\bf OIB}] \textit{One-request, Intra-binary.} A single request is able to trigger the bug inside a network-facing process.

\item[{\bf MIB}] \textit{Multi-request, Intra-binary.} Several requests are needed to trigger the bug inside a network-facing process.

\item[{\bf OID}] \textit{One-request, Inter-binary Direct.} A single request is able to trigger the bug inside a child spawned by a network-facing process.

\item[{\bf OII}] \textit{One-request, Inter-binary Indirect.} A single request is able to trigger the bug inside a binary that is not spawned by the network-facing process.

\item[{\bf MID}] \textit{Multi-request, Inter-binary Direct.} Several requests are needed to trigger the bug inside a child spawned by a network-facing process.

\item[{\bf MII}] \textit{Multi-request, Inter-binary Indirect.} Several requests are needed to trigger the bug inside a binary that is not spawned by the network-facing process.

\end{itemize}

These categories differ with respect to the number of requests needed to trigger the bug and where the bug occurs (the network-facing process, a child of the network-facing process, or another unrelated process). Fuzzers like TriforceAFL implicitly focus on OIB and OID bugs, where the crash is exercised by a single request and primarily affects the network-facing process (or its children). Protocol-aware fuzzers like AFLNet extend their attention to MIB bugs, which are triggered by multiple requests and affect the network-facing process. Our system-mode port of AFLNet, AFLNet$^{\dagger}$, should nonetheless identify MID and MII bugs. \ourtool{} naturally considers all the categories but embraces design choices that focus on bugs triggered by multiple requests, regardless of the crashing process.

The per-bug breakdown in Table~\ref{table:tte_bugs} confirms the fuzzers' nature. TriforceAFL can indeed find only OIB and OID bugs, with a TTE that can be for a few bugs even smaller than what is recorded for AFLNet$^{\dagger}$ and \ourtool{}. This is consistent with its lack of multi-request reasoning and single-process fuzzing scope. Overall, it finds 8 OIB bugs and 1 OID bug. However, it still misses 3 OIB bugs and 1 OID bugs that are found by \ourtool{}, likely due to the lack of taint-guided mutations. Interestingly, TriforceAFL quickly finds one OIB bug in TL-WPA8630 that is missed by the other two fuzzers. No OII bugs have been identified by \ourtool{},  TriforceAFL and AFLNet$^{\dagger}$. 

AFLNet$^{\dagger}$ is significantly more effective than TriforceAFL when considering bugs requiring multiple requests. Indeed, it can find 6 OIB bugs, 2 OID bugs, 2 MIB bugs, 15 MID bugs, and 3 MII bugs. This substantial improvement is likely due to being protocol-aware, which allows it to perform mutations on the multiple requests composing the interactions, preserving the protocol state.

\ourtool{} further improves on AFLNet$^{\dagger}$, discovering 10 OIB bugs, 3 OID bugs, 1 MIB bug, 21 MID bugs, and 7 MII bugs. The improvement likely results from the effective combination of state-aware mutations and taint-guided mutations. Moreover, not only does it find more unique bugs, but it can find them more consistently across runs. Indeed, for several bugs, \ourtool{} can find the bugs during most or all runs. However, there are still some bugs that it can find only within specific runs, highlighting the non-deterministic nature of fuzzing.

Finally, \ourtool{} maintains a nice balance between bug discovery effectiveness and TTE efficiency since it finds significantly more bugs than the competitors and for most bugs exhibits on average comparable TTEs.

\begin{answer}{2}
\ourtool{} can find 367\% and 50\% more unique bugs than TriforceAFL and AFLNet$^{\dagger}$, respectively. This improvement does not come at the cost of efficiency, since, on average, it exhibits a comparable TTE. Finally, \ourtool{} is more consistent than AFLNet$^{\dagger}$ across runs at finding the bugs, making it a more reliable framework solution.
\end{answer}


\subsection{RQ3: Bug Case Studies}
\label{ssec:rq3}

In this section, we analyze the bugs found by \ourtool{}. For the sake of brevity, we focus on the OID, MID, and MII bug categories, which are the categories not explicitly considered by the state-of-the-art of firmware fuzzing and that were central in inspiring the design of \ourtool{}. For each category, we provide one case study.



\subsubsection{[OID] TV-IP121WN: stack BOF in network.cgi}
\label{sssec:trendnet_networkcgi_ipcat}

The OID bug was found in the firmware image of the TRENDnet IP camera TV-IP121WN. The crash is triggered by a single HTTP POST processed by the webserver \texttt{boa} and then handled by the CGI binary \texttt{network.cgi}, thus it was categorized as an \emph{inter-binary} issue ({\tt boa} $\rightarrow$ {\tt network.cgi}). In particular, the attacker can craft a malformed POST value \texttt{ip4}, which after being processed but not sanitized by \texttt{boa} and passed to \texttt{network.cgi}, results in a stack buffer overflow within the function \texttt{ipCat()}, corrupting the return address and thus generating a {\tt SIGSEGV}. Since \texttt{boa} respawns the CGI on exit, repeated POSTs can repeatedly crash the CGI process, yielding a reliable remote DoS. Unfortunately, the bug occurs even with unauthenticated requests: \texttt{network.cgi} crashes even without valid session credentials because authentication is enforced after CGI invocation.

In more detail, the unauthenticated POST request is:
\begin{enumerate}[label=\protect\circnum{\arabic*}]
\item {\em Poisoned IP value.}  
  The attacker issues an HTTP POST to \texttt{/admin/network.cgi} where \texttt{ip1}, \texttt{ip2}, \texttt{ip3} can be benign values but \texttt{ip4} is a large attacker-controlled string. An example of the request is:
{\small
\begin{Verbatim}[commandchars=\\\{\}]
POST /admin/network.cgi HTTP/1.1
Host: 127.0.0.1
Content-Type: application/x-www-form-urlencoded
Content-Length: <...>
iptype=0&ip1=192&ip2=168&ip3=10&ip4=\colorbox{black}{\textcolor{white}{[>30 bytes attacker-}}
\colorbox{black}{\textcolor{white}{controlled string]}}&...other fields...
\end{Verbatim}
}
The webserver \texttt{boa} performs repeated \texttt{strlen}, \texttt{malloc}, and \texttt{memcpy} operations to build CGI inputs and then directly executes in a child the CGI process \texttt{network.cgi}. The \texttt{main} function from \texttt{network.cgi} loops over the form fields (executing several \texttt{strcmp} calls), records indices into an array, and then calls \texttt{ipCat()} to assemble dotted-IP strings into stack buffers:
{\small
\begin{verbatim}
  /* stack locals in decompilation */
  undefined auStack_348 [32];   /* destination buffer */
  ...
  /* assembly of dotted IPs (called after parsing) */
  ipCat(auStack_348,0x1e, ip1_ptr, ip2_ptr, ip3_ptr, ip4_ptr);
\end{verbatim}
}\vspace{-2mm}
The destination length argument is \texttt{0x1e} (30) while each octet pointer is taken verbatim from the CGI input prepared by \texttt{boa}; no per-octet length or character-class checks occur prior to the call.
\texttt{ipCat()} uses unbounded formatting/concatenation (for example, \texttt{sprintf(dst, "\%s.\%s.\%s.\%s", ...)}) and can therefore write arbitrarily long strings into the small \texttt{dst}. A long \texttt{ip4} component overruns \texttt{auStack\_348}, corrupting adjacent stack data (including the saved return address), and causes a crash when control returns.
\end{enumerate}

\subsubsection{[MID] DAP-2310: BOF in atp (redirect\_htm)}
\label{sssec:dap2310_xgi_bug}

The MID bug was found in the firmware image of the D-Link DAP-2310 router. The crash occurs inside the web/CGI handling process \texttt{xgi} but requires a multi-stage request flow: a web page emits an unescaped query, the CGI parser in the native \texttt{atp} module accumulates parameters without bounds checks, and finally \texttt{redirect\_htm} formats an HTTP header using unsafe string functions. The exploit is deterministic and proceeds in two steps:

\begin{enumerate}[label=\protect\circnum{\arabic*}]
\item {\em Authentication.}
An initial login POST is required to obtain a session, as in the motivating example bug (Section~\ref{sec:motivation}).
\item {\em Client-side redirect construction.}
The attacker can cause the browser to issue a GET request to \texttt{cfg\_valid} \texttt{.xgi} with query parameters of unbounded length. An example of the request is:
{\small
\begin{Verbatim}[commandchars=\\\{\}]
GET /cfg_valid.xgi?random=\colorbox{black}{\textcolor{white}{[>256 bytes]}}
&exeshell=submit%20COMMIT&exeshell=\colorbox{black}{\textcolor{white}{[>256 bytes]}}
%20WLAN&flag=1 HTTP/1.1
\end{Verbatim}
}\vspace{-2mm}
Both the \texttt{random} parameter and each \texttt{exeshell} argument may contain arbitrarily long attacker input. These strings are injected verbatim by \texttt{cfg\_valid.php}, which generates the redirect without encoding or length checks:
{\small
\begin{verbatim}
var str="cfg_valid.xgi?random="+generate_random_str();
function exe_str(str_shellPath) {
  myShell = str_shellPath.split(";");
  for(i=0; i<myShell.length; i++) {
    temp_str+="&"+"exeshell="+myShell[i];
  }
  return temp_str;
}
...
echo "str+=exe_str(\"submit COMMIT;\".$SUBMIT_STR.\"\");\n";
\end{verbatim}
}

Because the values are inserted unescaped, the CGI parser receives an oversized request. During parsing, the \texttt{atp} module copies names and values into fixed buffers using unsafe functions. Finally, \texttt{redirect\_htm} builds an HTTP 302 response header using two adjacent stack buffers:
{\small
\begin{verbatim}
char acStack_37c[100];
char local_318;
/* 256-byte "value buffer" */
undefined auStack_317[255];
char local_218;
/* 516-byte "header buffer" */
undefined auStack_217[515];
strcat(&local_218,"HTTP/1.0 302 Found\n");
/* Overflow */
sprintf(&local_318,"Location: %s\r\n", param_1);
strcat(&local_218,&local_318);
fputs(&local_218,stdout);
\end{verbatim}
}

The 256-byte value buffer is the first overflow point: a long \texttt{param\_1} string from the query (e.g., oversized \texttt{random} or \texttt{exeshell}) causes \texttt{sprintf} to overrun it. The 516-byte header buffer is subsequently at risk when concatenating the expanded \texttt{Location:} line; once the value buffer is corrupted, the header assembly may propagate the corruption further. Thus the attacker controls the crash via unbounded query parameters that flow from PHP into the parser and finally into \texttt{redirect\_htm}.
\end{enumerate}

\subsubsection{[MII] DIR-300: NULL dereference in udhcpd}
\label{sssec:dir300_udhcpd_bug}

The MII vulnerability that we consider was found in the firmware images of the D-Link DIR-300. The chain of this bug spans several processes: the HTTP server, the CGI handler and the DHCP daemon. In particular, the crashing DHCP process \texttt{udhcpd} has no direct process relationship with the HTTP daemon \texttt{httpd}, since it was as a daemon from the \texttt{init} process. The bug exploit requires a four-step sequence:


\begin{enumerate}[label=\protect\circnum{\arabic*}]
\item {\em Authentication.} The attacker first obtains a valid web session by logging in:
{\small
\begin{verbatim}
POST /login.php HTTP/1.1
Host: 192.168.0.1
Content-Type: application/x-www-form-urlencoded

ACTION_POST=LOGIN&LOGIN_USER=admin&LOGIN_PASSWD=&login=
Log+In
\end{verbatim}
}

\item {\em Poisoned in-memory runtime configuration.} Using the authenticated session, the attacker issues a POST that writes a set of data form fields into an in-memory runtime configuration stored under the {\tt tmpfs}-mounted {\tt /runtime}. On the DIR-300 this is handled by \texttt{set\_temp\_nodes.php}, which stores each \texttt{d\_<i>\_<j>} form field verbatim under paths such as \texttt{/runtime/post/session\_\allowbreak<sid>/entry:<i>/data\_<j>}. This files likely represent the internal XML database used to store the  configuration. Example of the request (truncated):
{\small
\begin{Verbatim}[commandchars=\\\{\}]
POST /set_temp_nodes.php HTTP/1.1
...
TEMP_NODES=/runtime/post/session_1&data=4&start=1
&d_1_2=192.168.0.45&d_1_3=ce%3A92%3Abf%3Ad1%3A09%3Acf
...
&d_3_2=192.168.0.3&d_3_3=\colorbox{black}{\textcolor{white}{[attacker-controlled string]}}
&end=25
\end{Verbatim}
}

The script stores the posted values without server-side sanitization, so a crafted or overly long token injected here becomes persisted in the runtime configuration.

\item {\em Poisoned persistent configuration.} When the LAN/DHCP form is submitted through the request (simplified):
{\small
\begin{verbatim}
POST /bsc_lan.php HTTP/1.1
Host: 192.168.0.1
Content-Type: application/x-www-form-urlencoded

ACTION_POST=SOMETHING&...&startipaddr=192.168.0.100&
endipaddr=192.168.0.199&...
\end{verbatim}
}\vspace{-2mm}

the execution of the \texttt{bsc\_lan.php} handler is triggered, which copies the runtime configuration into the persistent configuration files under the NVRAM path {\tt /lan} (e.g., \texttt{/lan/dhcp/}\allowbreak{\tt server/pool:1/staticdhcp/entry:<N>/ }\allowbreak{\tt \{hostname, ip, mac\}}) via the {\tt set} function. Unfortunately, \texttt{bsc\_lan.php} copies values read from \texttt{/runtime/}\allowbreak{\tt post/...} into the \texttt{/lan/...} files without further validation, so any poisoned string from step (2) is now part of the device configuration.

\item {\em Configuration commit and crash.} Finally, a GET request to \texttt{bsc\_lan.xgi} issues the commit of the configuration through \texttt{exeshell=submit\allowbreak\%20COMMIT} and restarts the DHCP service through \texttt{exeshell=submit\allowbreak\%20DHCPD}:
{\small
\begin{verbatim}
GET /bsc_lan.xgi?random_num=...&exeshell=submit%20COMMIT
&exeshell=submit%20DHCPD HTTP/1.1
Host: 192.168.0.1
\end{verbatim}
}

This script triggers the execution of the template engine \texttt{/etc/templates/dhcp/dhcpd.php}, which reads the persistent \texttt{/lan/dhcp/...} files and writes a new \texttt{udhcpd} configuration file \texttt{/var/run/udhcpd-br0.conf} using unchecked \texttt{fwrite}/\texttt{fwrite2} calls, then restarts the DHCP daemon \texttt{udhcpd}.

When \texttt{udhcpd} parses the poisoned file, it calls its configuration reader (\texttt{read\_config}) which extracts key-value pairs from each line. For certain malformed or empty values in the configuration file, the pointer to the value returned by the internal lookup function can be \texttt{NULL}. This pointer is then passed directly to a resolver wrapper (decompiled as \texttt{FUN\_00402f70}) and into \texttt{inet\_aton(param\_1, ...)} without any check for \texttt{NULL}, producing a NULL-pointer dereference and causing a {\tt SIGSEGV}:
{\small
\begin{verbatim}
undefined4 FUN_00402f70(char *param_1, in_addr *param_2)
{
  int i = inet_aton(param_1, param_2); /* NO NULL CHECK */
  ...
}
\end{verbatim}
}
\end{enumerate}

\begin{answer}{3}
The bug case studies considered in this section demonstrate that \ourtool{} successfully discovers interesting and non-trivial vulnerabilities arising from multi-request sequences and inter-binary interactions, confirming that our approach achieves the initial goals outlined in this work.
\end{answer}

\subsection{RQ4: Taint Hint Causality}

In this section, we investigate \emph{why} \ourtool{} was able to discover the bugs discussed in the previous sections. In particular, we analyze the role of taint hints to understand their impact during fuzzing by defining the \emph{causality score}. For each crashing input, we assign a causality score of $1.0$ if the input results from at least one taint-guided mutation in the chain of mutations that transformed the original seed into the crashing input, and $0$ otherwise. We then compute the average across different runs.


\paragraph{Aggregate analysis.}
Among all 42 bugs found by \ourtool{}, we found that 66.7\% (28) of bugs have full taint causality (score 1.0), indicating they are consistently discovered, even across different runs, through taint-guided mutations. Another 9.5\% (4) show high causality (scores 0.5--0.99), 7.1\% (3) show low causality (scores 0.1--0.49), and 16.7\% (7) show no causality (score $<0.1$), suggesting these bugs could be discovered through state-aware mutations alone.

Table~\ref{tab:taint_causality_summary} shows the average causality score when grouping bugs by category. The OIB, OID, and MID categories exhibit high causality, demonstrating that taint analysis is particularly crucial for discovering inter-binary bugs that require precise dependency tracking. The lower MII score reflects cases where complex multi-process chains can sometimes be triggered through state-aware mutations. Finally, the MIB category shows no causality but represents only a single bug instance.

\begin{table}[t]
\centering
\renewcommand{\arraystretch}{1.2}
\setlength{\tabcolsep}{8pt}
\begin{tabular}{|l|c|c|}
\hline
{\sc Bug Category} & {\sc \# Found Bugs} & {\sc Causality Score} \\
\hline\hline
OIB & 10 & 0.75 \\
OID & 3 & 0.8 \\
MIB & 1 & 0.0 \\
MID & 21 & 0.83 \\
MII  & 7 & 0.57 \\
\hline
\textbf{ALL} & \textbf{42} & \textbf{0.79} \\
\hline
\end{tabular}
\caption{Causality scores per bug category.}
\label{tab:taint_causality_summary}
\end{table}

\paragraph{Case studies}
We now analyze the crashing inputs for the three case studies from RQ3 to understand the precise mechanisms by which taint hints enable bug discovery, examining mutation patterns across all runs for each vulnerability.

For the OID bug from TV-IP121WN, we measured a full causality score of 1.0. The stack buffer overflow demonstrates full causality with consistent targeting of the \texttt{ip4} parameter across all 5 runs. In 4 runs, taint hints targeted offset 889 (the \texttt{3} in \texttt{ip4=30}), while in 1 run they targeted offset 890 (the \texttt{0}). Both mutations inserted 1260-byte havoc strings into the IP parameter, triggering the overflow in \texttt{ipCat()}. The dependency analysis consistently identified only vulnerable region 18 as required, filtering out all 21 other regions. This precision demonstrates how taint analysis enables surgical targeting of the exact parameter bytes that flow to the vulnerable function, explaining why random mutation of the 399-byte POST body would have minimal success probability.

For the MID bugs from DAP-2310, we also observed a full causality score of 1.0. Full causality manifests through diverse but consistent targeting of query parameters across all 5 runs. The taint hints targeted different offsets within the \texttt{exeshell} parameter: offset 113 targeting \texttt{DELAY\_LAN}, offset 64 and 116 targeting \texttt{COMMIT}, and offset 69 targeting the parameter name itself. Crucially, one run (region 34, offset 69) achieved the optimal dependency minimization described in RQ3, filtering out 11 of 14 regions to retain only the essential causal chain. The other runs filtered out 7 regions, still providing significant sequence reduction. This variability shows taint analysis identifying multiple vulnerable injection points within the same attack vector while consistently maintaining the multi-request dependencies necessary for exploitation.

For the MII bugs from DIR-300, we also obtained  a full causality score of 1.0. Full causality across all 5 runs shows systematic targeting of MAC address fields in the configuration data (region 32). Taint hints consistently targeted URL-encoded MAC addresses: offsets 679 and 680 in {\tt d\_3\_3} field, and offset 604 in {\tt d\_2\_3} field, applying single-byte bitflip mutations. All runs achieved optimal dependency minimization, filtering out 69 of 79 regions to preserve only the 10-region causal chain spanning authentication, configuration poisoning, and commit operations. The precision of targeting specific bytes within hex-encoded MAC addresses (e.g., changing a byte to an invalid one) while maintaining the complex multi-stage dependency chain demonstrates taint analysis' capability to track data flows across the complete 4-step attack sequence.


\begin{answer}{4}
Taint-guided mutations are responsible for discovering 83.3\% of bugs found by \ourtool{}, demonstrating highest effectiveness for multi-request, inter-binary vulnerabilities. \ourtool{}'s effectiveness stems from systematic identification of relevant input bytes and inter-region dependencies, enabling surgical precision in targeting vulnerable parameters while maintaining necessary multi-request chains.
\end{answer}

\section{Related Work}

Research related to \ourtool{} spans three main areas: protocol-aware and stateful fuzzing, firmware emulation and fuzzing, and dataflow-guided analysis for multi-step vulnerabilities. We briefly summarize representative prior work in each area and clarify how \ourtool{} differs.

\subsection{Protocol-Aware and Stateful Fuzzers}
Early protocol fuzzers, such as Sulley~\cite{amini2013sulley} or BooFuzz~\cite{pereyda2015boofuzz}, relied on manually written grammars or templates. AFLNet~\cite{pham2020aflnet} introduced coverage- and state-feedback for network protocols by replaying and mutating recorded message sequences and using server responses to guide exploration. StateAFL~\cite{natella2022stateafl} extends this idea by hashing memory snapshots to infer protocol states. Recent embedded- and web-focused tools improve the quality of inputs by exploiting request correlations or semantics: CinfoFuzz~\cite{feng2022cinfofuzz} leverages web-service correlation information for embedded web interfaces, SRFuzzer~\cite{zhang2019srfuzzer} applies semantic models and automated recovery to test physical SOHO routers, and app-driven or blackbox approaches, such as IoTFuzzer~\cite{chen2018iotfuzzer}, Diane~\cite{redini2021diane}, and Snipuzz~\cite{redini2021diane}, use companion apps or response-based inference to produce higher-quality test cases without firmware access.

These systems improve per-service state exploration and input validity, but typically operate on a single daemon or device channel. \ourtool{} generalizes protocol-aware ideas to whole, rehosted Linux firmware by discovering dependencies across services (network, filesystem, IPC) and synthesizing multi-request, cross-component sequences for stateful fuzzing.

\subsection{Firmware Emulation and Fuzzing}
Large-scale firmware testing depends on robust rehosting. Firmadyne~\cite{chen2016towards} pioneered automated extraction and system-mode emulation for Linux firmware; FirmAE~\cite{kim2020firmae} boosted emulation success through arbitration heuristics. TriforceAFL~\cite{triforceafl}, FirmFuzz~\cite{srivastava2019firmfuzz}, and related projects integrate fuzzing into QEMU-based environments to allow full-system testing. Throughput-oriented efforts -- FIRM-AFL~\cite{zheng2019firm}, EQUAFL~\cite{equafl}, SAFIREFUZZ~\cite{seidel2023forming} -- explore hybrid or near-native execution to reduce emulator overhead and speed fuzzing campaigns. For MCU or bare-metal firmware, automated peripheral-modeling/re-hosting -- Laelaps~\cite{cao2020device}, P2IM~\cite{feng2020p2im}, Jetset~\cite{johnson2021jetset}, µEmu~\cite{zhou2021automatic}, HALucinator~\cite{clements2020halucinator}, Fuzzware~\cite{scharnowski2022fuzzware} -- infer or synthesize peripheral behavior to make firmware executable without per-device engineering. Firmulti~\cite{cheng2023firmulti} demonstrates system-level VMI monitoring for multi-process firmware analysis.

These works solve the orthogonal problems of making firmware executable and doing so at scale or high throughput. \ourtool{} assumes modern Linux-based firmware can be rehosted using these advances and focuses on exercising the interactions among co-running services by combining rehosting with whole-system taint-driven dependency discovery and stateful fuzzing.

\subsection{Dataflow-Guided Fuzzing}
Process-level taint- and dataflow-guided fuzzers, such as VUzzer~\cite{rawat2017vuzzer}, Angora~\cite{chen2018angora}, RedQueen~\cite{aschermann2019redqueen}, markedly improve the ability to bypass complex checks by identifying influential input bytes. Whole-system taint engines such as DECAF++~\cite{davanian2019decaf++} enable VM-level tracking across kernel and userspace, while static multi-binary analyses -- Karonte~\cite{redini2020karonte}, SaTC~\cite{chen2023satc} -- surface insecure inter-binary flows in firmware. Recent work stresses the importance of multi-step, cross-component vulnerabilities: HOFF~\cite{yu2021towards} detects higher-order memory corruption by tracking data flows from attacker-controlled stores to memory-critical operations, and ReLink~\cite{yu2021relink} uncovers higher-order command injection by linking requests across interfaces. Complementary to taint analysis, lightweight runtime sanitizers -- e.g., DFirmSan~\cite{yang2025dfirmsan} -- aim to increase observable fault signals in firmware fuzzing with reduced overhead.

Unlike single-process taint-guided fuzzers or tools that only detect multi-step issues, \ourtool{} integrates whole-system taint into an active fuzzing loop: it uses cross-component taint to
\begin{enumerate*}[label=(\roman*)]
\item identify which bytes and requests influence code regions,
\item prioritize mutations accordingly, and
\item construct minimal, checkpointed multi-request sequences that exercise multi-step behaviors across daemons.
\end{enumerate*}


\section{Limitations}
\label{sec:limitations}

While \ourtool{} demonstrates significant improvements in firmware fuzzing effectiveness, several limitations constrain its applicability and scope.

\paragraph{Limited Device Scope.} Our evaluation focuses on MIPS-architecture Linux-based firmware from routers, range extenders, and IP cameras. This represents only a subset of the embedded device ecosystem, excluding ARM/x86 architectures, RTOS/bare-metal systems, and other device categories. The generalizability to modern firmware with security features like encryption remains unvalidated.

\paragraph{Emulation and Protocol Dependencies.} \ourtool{} inherits the fundamental challenges of firmware rehosting, limiting applicability to devices requiring specialized peripherals or precise timing. The current HTTP-focused implementation excludes devices using other protocols (FTP, DHCP, SMTP, etc.), while manual seed corpus generation limits scalability for large-scale analysis.

\paragraph{Taint Analysis Approximations.} The byte annotation heuristic achieves good precision but represents an approximation of full per-byte taint tracking. The single taint source approach may miss complex transformations or generate false positives when similar patterns appear in unrelated contexts, requiring careful interpretation of taint hints during mutation targeting.

\paragraph{Computational Overhead.} Despite optimizations, whole-system taint analysis introduces substantial overhead compared to application-level fuzzing, potentially limiting adoption in resource-constrained environments. The dependency analysis phase requires significant computational resources for large firmware images.

\section{Conclusions}
\label{sec:conclusions}

This article presents \ourtool{}, a fuzzing framework aimed at firmware images exposing HTTP-based services that control and interact with multiple cooperating daemons. \ourtool{} analyzes the firmware execution to generate \emph{taint hints}, which represent the influence of bytes from specific network requests on specific code portions within the firmware. Taint hints are used to guide the fuzzing by focusing mutations on relevant byte subsequences and replaying only a reduced subset of the original requests necessary to preserve execution context.

An experimental evaluation across a diverse dataset of real-world firmware images shows that \ourtool{} consistently outperforms existing state-of-the-art firmware fuzzing frameworks in both the number and reproducibility of discovered bugs. In particular, \ourtool{} can identify bugs requiring complex sequences of network requests involving different firmware daemons. We investigated several case studies to demonstrate that the discovered bugs indeed depend on the taint hints constructed by \ourtool{}.




\section*{Acknowledgments}
This work was partially supported by:
\begin{itemize}
  \item Project FARE (PNRR M4.C2.1.1 PRIN 2022, Cod. 202225BZJC, CUP D53D23008380006, Avviso D.D 104 02.02.2022) under the Italian NRRP MUR program funded by the European Union - NextGenerationEU.
  
  \item Project SETA (PNRR M4.C2.1.1 PRIN 2022 PNRR, Cod. P202233M9Z, CUP B53D23026000001, Avviso D.D 1409 14.09.2022) under the Italian NRRP MUR program funded by the European Union - NextGenerationEU.

  \item Project SERICS (PE00000014) under the NRRP MUR program funded by the EU-NGEU
\end{itemize}

\balance
\bibliographystyle{cas-model2-names}



\end{document}